\documentclass[10pt,conference]{IEEEtran}

\usepackage{xspace}
\usepackage{xcolor}
\usepackage{graphicx}
\usepackage[normalem]{ulem} %
\usepackage{amsmath}
\usepackage{amssymb}
\usepackage{amsfonts}
\usepackage{enumitem}
\usepackage{textcomp}
\usepackage{xstring}

\usepackage[linkcolor=black,citecolor=black,anchorcolor=black,filecolor=black,menucolor=black,runcolor=black,urlcolor=black,hidelinks]{hyperref}
\usepackage{breakurl}
\usepackage{cite}

\usepackage{booktabs}
\usepackage{multirow}
\usepackage{makecell}
\usepackage{ragged2e}

\usepackage{caption}
\usepackage{subcaption}

\usepackage{listings}

\usepackage{tikz}
\usetikzlibrary{calc}
\usetikzlibrary{shapes.geometric}
\usetikzlibrary{decorations.pathreplacing}
\usetikzlibrary{positioning}
\usetikzlibrary{automata}
\usetikzlibrary{arrows}

\usepackage{datetime} %

\newcommand{\XSpace}[1]{}
\newcommand{\XComment}[1]{}

\newcommand{\DefMacro}[2]{\expandafter\newcommand\csname rmk-#1\endcsname{#2}}
\newcommand{\UseMacro}[1]{\csname rmk-#1\endcsname}

\newcommand{\MyPara}[1]{\vspace{1pt}\noindent\textbf{#1}.}

\newcommand{\InputWithSpace}[1]{\bgroup\def\arraystretch{1.1}\input{#1}\egroup}
\newcommand{\Code}[1]{{\ifmmode{\mathtt{#1}}\else$\mathtt{#1}$\fi}}
\newcommand{\CodeIn}[1]{{\ifmmode{\mathtt{#1}}\else$\mathtt{#1}$\fi}}
\newcommand{\CoqIn}[1]{\lstinline[language=Coq,basicstyle=\normalsize\ttfamily]{#1}}

\newcolumntype{R}[1]{>{\RaggedLeft\arraybackslash}p{#1}}
\newcolumntype{L}[1]{>{\RaggedRight\arraybackslash}p{#1}}

\newcommand{\numtoword}[1]{%
\IfStrEqCase{#1}{{0}{zero}{1}{one}{2}{two}{3}{three}{4}{four}{5}{five}{6}{six}{7}{seven}{8}{eight}{9}{nine}{10}{ten}}[#1]}

\definecolor{gray}{RGB}{211,211,211}
\newcommand{\jbasicstyle}{\small\sffamily} %

\newcommand{\jnumberstyle}{\scriptsize}

\lstdefinelanguage{pseudo}
{
  morekeywords={},
  keywordstyle=\bfseries,
  lineskip=-0.1em,
  numbers=left, %
  numberstyle=\jnumberstyle,
  numbersep=4pt,
  basicstyle=\jbasicstyle,
  breaklines=true,
  breakautoindent=true,
  tabsize=2,
  columns=fullflexible,
  morecomment=*[l][\textsl]{//},
  mathescape=true,
  xleftmargin=10pt,
}

\lstdefinelanguage{todo-comment}
{
  morekeywords={},
  keywordstyle=\bfseries,
  lineskip=-0.1em,
  numbers=none,
  basicstyle=\jbasicstyle,
  breaklines=true,
  breakautoindent=true,
  tabsize=2,
  columns=fullflexible,
  morecomment=*[l][\textsl]{//},
  mathescape=true,
  xleftmargin=-10pt,
}

\lstdefinelanguage{java-pretty}
{
  language=java,
  numbers=left,
  basicstyle=\scriptsize\ttfamily,
  numberstyle=\scriptsize,
  breaklines=true,
  columns=fullflexible,
  xleftmargin=16pt,
  showstringspaces=false,
}

\lstdefinelanguage{cmdline}
{
  language=pseudo,
  numbers=none,
  basicstyle=\footnotesize\ttfamily,
  numberstyle=\scriptsize,
  breaklines=true,
  columns=fullflexible,
  xleftmargin=-2pt,
  xrightmargin=-2pt,
  showstringspaces=false,
}

\definecolor{shadecolor}{gray}{1.00}
\definecolor{darkgray}{gray}{0.30}
\definecolor{violet}{rgb}{0.56, 0.0, 1.0}
\definecolor{forestgreen}{rgb}{0.13, 0.55, 0.13}

\lstdefinelanguage{Coq} {
mathescape=true,						
texcl=false,
morekeywords=[1]{
  Add,
  All,
  Arguments,
  Axiom,
  Bind,
  Canonical,
  Check,
  Close,
  CoFixpoint,
  CoInductive,
  Coercion,
  Contextual,
  Corollary,
  Defined,
  Definition,
  Delimit,
  End,
  Example,
  Export,
  Fact,
  Fixpoint,
  Goal,
  Graph,
  Hint,
  Hypotheses,
  Hypothesis,
  Implicit,
  Implicits,
  Import,
  Inductive,
  Lemma,
  Let,
  Local,
  Locate,
  Ltac,
  Maximal
  Module,
  Morphism,
  Next,
  Notation,
  Obligation,
  Open,
  Parameter,
  Parameters,
  Prenex,
  Print,
  Printing,
  Program,
  Projections,
  Proof,
  Proposition,
  Qed,
  Record,
  Relation,
  Remark,
  Require,
  Reserved,
  Resolve,
  Rewrite,
  Save,
  Scope,
  Search,
  Section,
  Show,
  Strict,
  Structure,
  Tactic,
  Theorem,
  Unset,
  Variable,
  Variables,
  View,
  inside,
  outside
},
morekeywords=[2]{
  as,
  cofix,
  else,
  end,
  exists,
  exists2,
  fix,
  for,
  forall,
  fun,
  if,
  in,
  is,
  let,
  match,
  nosimpl,
  of,
  return,
  struct,
  then,
  vfun,
  with
},
morekeywords=[3]{Type, Prop, Set, True, False},
morekeywords=[4]{
  after,
  apply,
  assert,
  auto,
  bool_congr,
  case,
  change,
  clear,
  compute,
  congr,
  cut,
  cutrewrite,
  destruct,
  elim,
  field,
  fold,
  generalize,
  have,
  heval, 
  hnf,
  induction,
  injection,
  intro,
  intros,
  intuition,
  inversion,
  left,
  loss,
  move,
  nat_congr,
  nat_norm,
  pattern,
  pose,
  refine,
  rename,
  replace,
  revert,
  rewrite,
  right,
  ring,
  set,
  simpl,
  split,
  subst,
  suff,
  suffices,
  symmetry,
  transitivity,
  trivial,
  unfold,
  unlock,
  using,
  without,
  wlog,
  autorewrite
},        
morekeywords=[5]{
  assumption,
  by,
  contradiction,
  congruence,
  done,
  exact,
  lia,
  gappa,
  omega,
  reflexivity,
  romega,
  solve,
  tauto,
  discriminate,
  unsat
},
morecomment=[s]{(*}{*)},
morekeywords=[6]{do, first, try, idtac, repeat},
showstringspaces=false,
morestring=[b]",
tabsize=3,							
extendedchars=true,  		 		
sensitive=true, 
breaklines=false,
basicstyle=\footnotesize\ttfamily,
captionpos=b,							
columns=[l]fullflexible,
identifierstyle={\color{black}},
keywordstyle=[1]{\color{violet}},
keywordstyle=[2]{\color{forestgreen}},
keywordstyle=[3]{\color{forestgreen}},
keywordstyle=[4]{\color{blue}},
keywordstyle=[5]{\color{red}},
keywordstyle=[6]{\color{violet}},
stringstyle=,
commentstyle=\it\ttfamily\color{brown},
numberstyle=\tiny%
}

\lstdefinestyle{Coq}{language=Coq}
\definecolor{highlightcolor}{RGB}{134,46,60}
\newcommand{\Light}[1]{\textcolor{highlightcolor}{\textbf{#1}}}
\newcommand*\circled[1]{\tikz[baseline=(char.base)]{
            \node[shape=circle, draw=black, minimum size=0.2, inner sep=2pt, fill=white, thick, text=black] (char) {#1};}}

\newcommand{\CoqConvTool}{\textsc{Roosterize}\xspace}
\newcommand{\Title}{\CoqConvTool: Suggesting Lemma Names for Coq Verification Projects Using Deep Learning}

\newcommand{\LemmaN}{Lemma Names\xspace}
\newcommand{\lemman}{lemma names\xspace}

\newcommand{\Coq}{Coq\xspace}
\newcommand{\tooltype}{tool\-chain\xspace}
\newcommand{\Tooltype}{Tool\-chain\xspace}

\newcommand{\cli}{command-line interface\xspace}

\newcommand{\DSta}{All Tiers\xspace}

\newcommand{\DSti}{Tier 1\xspace}

\newcommand{\DStii}{Tier 2\xspace}

\newcommand{\DStiii}{Tier 3\xspace}
\newcommand{\encdec}{encoder-decoder\xspace}

\newcommand{\ktree}{kernel tree\xspace}
\newcommand{\ktrees}{kernel trees\xspace}
\newcommand{\Ktrees}{Kernel trees\xspace}

\newcommand{\KTreeAcro}{KnlTree\xspace}

\newcommand{\lstmt}{\lemmastmt{}\xspace}
\newcommand{\Lstmt}{\Lemmastmt{}\xspace}

\newcommand{\LeftOutCorpus}{Left-out\xspace}
\newcommand{\lemmastmt}{lemma statement\xspace}
\newcommand{\lemmastmts}{lemma statements\xspace}
\newcommand{\Lemmastmt}{Lemma statement\xspace}

\newcommand{\LStmtAcro}{Stmt\xspace}

\newcommand{\Ngram}{N-gram\xspace}
\newcommand{\nnmodel}{neural network model\xspace}
\newcommand{\NNModel}{Neural Network Model\xspace}
\newcommand{\retrievalbased}{retrieval-based\xspace}
\newcommand{\RetrievalBased}{Retrieval-based\xspace}

\newcommand{\stree}{syntax tree\xspace}
\newcommand{\strees}{syntax trees\xspace}
\newcommand{\Strees}{Syntax trees\xspace}
\newcommand{\STreeAcro}{SynTree\xspace}

\newcommand{\sktrees}{syntax and kernel trees\xspace}

\newcommand{\subtok}{sub-token\xspace}

\newcommand{\subtokenize}{sub-tokenize\xspace}
\newcommand{\subtokenized}{sub-tokenized\xspace}
\newcommand{\subtokenizer}{sub-tokenizer\xspace}

\newcommand{\trimmed}{chopped\xspace} %
\newcommand{\TrimmedAcro}{Chop\xspace}
\newcommand{\trimmedktree}{\trimmed \ktree}

\newcommand{\TrimmedKTreeAcro}{\TrimmedAcro{}\KTreeAcro{}\xspace}
\newcommand{\trimmedstree}{\trimmed \stree}

\newcommand{\TrimmedSTreeAcro}{\TrimmedAcro{}\STreeAcro{}\xspace}
\newcommand{\trimming}{chopping\xspace}

\newcommand{\MathComp}{MathComp\xspace}

\newcommand{\seqtoseq}{\textsc{seq2seq}\xspace}
\newcommand{\SerAPI}{SerAPI\xspace}

\newcommand{\VSCode}{VSCode\xspace}

\newcommand{\sertok}{\texttt{sertok}\xspace}
\newcommand{\sercomp}{\texttt{sercomp}\xspace}
\newcommand{\sername}{\texttt{sername}\xspace}

\newcommand{\doc}{file\xspace} %
\newcommand{\docs}{files\xspace} %

\newcommand{\tDataMiner}{\texttt{DataMiner}\xspace}
\newcommand{\tSubTokenizer}{\texttt{SubTok}\xspace}
\newcommand{\tModel}{\texttt{MISeq2Seq}\xspace}

\newcommand{\bleu}{BLEU\xspace}
\newcommand{\Bleu}{BLEU\xspace}
\newcommand{\fragacc}{fragment accuracy\xspace}

\newcommand{\toponeacc}{top-1 accuracy\xspace}

\newcommand{\topfiveacc}{top-5 accuracy\xspace}

\newcommand{\NGNLMF}{\Ngram LM\xspace}

\DefMacro{f-brnn}{BiRNNLM\xspace}
\DefMacro{f-brnn-CK}{BiRNNLM(C+K)\xspace}
\DefMacro{f-brnn-CS}{BiRNNLM(C+S)\xspace}
\DefMacro{f-brnn-SK}{BiRNNLM(S+K)\xspace}
\DefMacro{f-brnn-C}{BiRNNLM(C)\xspace}
\DefMacro{f-brnn-S}{BiRNNLM(S)\xspace}
\DefMacro{f-brnn-K}{BiRNNLM(K)\xspace}
\DefMacro{f-rrnn}{Reversed RNNLM\xspace}
\DefMacro{f-rnn}{RNNLM\xspace}
\DefMacro{f-ngram}{\NGNLMF{}\xspace}

\DefMacro{f-brnn+cl+rr}{BiRNNLM+CL+Reranking\xspace}
\DefMacro{f-brnn+masking+rr}{BiRNNLM+Masking+Reranking\xspace}
\DefMacro{f-ngram+rr}{\NGNLMF{}+Reranking\xspace}
\DefMacro{f-brnn+cl}{BiRNNLM+CL\xspace}
\DefMacro{f-brnn+rr}{BiRNNLM+Reranking\xspace}
\DefMacro{f-brnn+masking}{BiRNNLM+Masking\xspace}

\DefMacro{ln-s+bsexpl1+fsexpl1+attn+copy}{\LStmtAcro{}\allowbreak+\TrimmedKTreeAcro{}\allowbreak+\TrimmedSTreeAcro{}\xspace}
\DefMacro{ln-s+bsexpl1+attn+copy}{\LStmtAcro{}\allowbreak+\TrimmedKTreeAcro{}\xspace}
\DefMacro{ln-s+fsexpl1+attn+copy}{\LStmtAcro{}\allowbreak+\TrimmedSTreeAcro{}\xspace}
\DefMacro{ln-bsexpl1+fsexpl1+attn+copy}{\TrimmedKTreeAcro{}\allowbreak+\TrimmedSTreeAcro{}\xspace}
\DefMacro{ln-s+attn+copy}{\LStmtAcro{}\xspace}

\DefMacro{ln-rb}{\RetrievalBased{}\xspace}

\DefMacro{group-t1}{\DSti}
\DefMacro{group-t2}{\DStii}
\DefMacro{group-t3}{\DStiii}
\DefMacro{group-ta}{\DSta}
\DefMacro{group-lo}{\LeftOutCorpus}
\DefMacro{group-lo-small}{LO}
\DefMacro{group-allgroup}{All}
\DefMacro{lo-train-t1}{0}
\DefMacro{lo-train-ta}{0}

\DefMacro{table-caption-results-ln-t1--t1--t1-main}{
    \label{tbl:results-ln-t1--t1--t1-main}Results of \CoqConvTool
    Models.}

\newcommand{\TableHeadModel}{\textbf{Model}\xspace}

\DefMacro{table-head-bleu}{\bleu\xspace}
\DefMacro{table-head-frag-acc}{Frag.Acc.\xspace}
\DefMacro{table-head-full-acc-top-1}{Top1\xspace}
\DefMacro{table-head-full-acc-top-5}{Top5\xspace}
\DefMacro{table-head-results-ln-BLEU-4}{\textbf{\UseMacro{table-head-bleu}}\xspace}
\DefMacro{table-head-results-ln-frag-acc}{\textbf{\UseMacro{table-head-frag-acc}}\xspace}
\DefMacro{table-head-results-ln-full-acc-top-1}{\textbf{\UseMacro{table-head-full-acc-top-1}}\xspace}
\DefMacro{table-head-results-ln-full-acc-top-5}{\textbf{\UseMacro{table-head-full-acc-top-5}}\xspace}

\DefMacro{vspace-results-ln-t1--t1--t1-main}{-2pt}

\DefMacro{corpus-mathcomp-full-name}{math-comp\_math-comp}
\DefMacro{corpus-mathcomp-short-name}{math-comp}
\DefMacro{corpus-mathcomp-url}{https://github.com/math-comp/math-comp}
\DefMacro{corpus-mathcomp-url-pretty}{math-comp/math-comp}
\DefMacro{corpus-mathcomp-sha}{748d716efb2f2f75946c8386e441ce1789806a39}
\DefMacro{corpus-mathcomp-sha-pretty}{748d716}
\DefMacro{corpus-mathcomp-tag}{748d716efb2f2f75946c8386e441ce1789806a39}
\DefMacro{corpus-mathcomp-tier}{1}
\DefMacro{corpus-fcslpcm-full-name}{imdea-software\_fcsl-pcm}
\DefMacro{corpus-fcslpcm-short-name}{fcsl-pcm}
\DefMacro{corpus-fcslpcm-url}{https://github.com/imdea-software/fcsl-pcm}
\DefMacro{corpus-fcslpcm-url-pretty}{imdea-software/fcsl-pcm}
\DefMacro{corpus-fcslpcm-sha}{eef45034808487fa3fa7dc360514c6734efc8d07}
\DefMacro{corpus-fcslpcm-sha-pretty}{eef4503}
\DefMacro{corpus-fcslpcm-tag}{eef45034808487fa3fa7dc360514c6734efc8d07}
\DefMacro{corpus-fcslpcm-tier}{3}
\DefMacro{corpus-disel-full-name}{DistributedComponents\_disel}
\DefMacro{corpus-disel-short-name}{disel}
\DefMacro{corpus-disel-url}{https://github.com/DistributedComponents/disel}
\DefMacro{corpus-disel-url-pretty}{DistributedComponents/disel}
\DefMacro{corpus-disel-sha}{e8aa80c486ec618888b6d0c801da7f61b6046daa}
\DefMacro{corpus-disel-sha-pretty}{e8aa80c}
\DefMacro{corpus-disel-tag}{e8aa80c486ec618888b6d0c801da7f61b6046daa}
\DefMacro{corpus-disel-tier}{3}
\DefMacro{corpus-reglang-full-name}{palmskog\_coq-reglang}
\DefMacro{corpus-reglang-short-name}{reglang}
\DefMacro{corpus-reglang-url}{https://github.com/palmskog/coq-reglang}
\DefMacro{corpus-reglang-url-pretty}{palmskog/coq-reglang}
\DefMacro{corpus-reglang-sha}{da333e01771f41debc3086a06e45d1393533d5c4}
\DefMacro{corpus-reglang-sha-pretty}{da333e0}
\DefMacro{corpus-reglang-tag}{da333e01771f41debc3086a06e45d1393533d5c4}
\DefMacro{corpus-reglang-tier}{3}
\DefMacro{corpus-compdecpdl-full-name}{palmskog\_comp-dec-pdl}
\DefMacro{corpus-compdecpdl-short-name}{comp-dec-pdl}
\DefMacro{corpus-compdecpdl-url}{https://github.com/palmskog/comp-dec-pdl}
\DefMacro{corpus-compdecpdl-url-pretty}{palmskog/comp-dec-pdl}
\DefMacro{corpus-compdecpdl-sha}{c1f813b8c97805e3c676cb559d0eab8381988a8a}
\DefMacro{corpus-compdecpdl-sha-pretty}{c1f813b}
\DefMacro{corpus-compdecpdl-tag}{c1f813b8c97805e3c676cb559d0eab8381988a8a}
\DefMacro{corpus-compdecpdl-tier}{3}
\DefMacro{corpus-oddorder-full-name}{math-comp\_odd-order}
\DefMacro{corpus-oddorder-short-name}{odd-order}
\DefMacro{corpus-oddorder-url}{https://github.com/math-comp/odd-order}
\DefMacro{corpus-oddorder-url-pretty}{math-comp/odd-order}
\DefMacro{corpus-oddorder-sha}{ca602a4638a9fe8ac30780095543d861f60fbfa0}
\DefMacro{corpus-oddorder-sha-pretty}{ca602a4}
\DefMacro{corpus-oddorder-tag}{ca602a4638a9fe8ac30780095543d861f60fbfa0}
\DefMacro{corpus-oddorder-tier}{1}
\DefMacro{corpus-fourcolor-full-name}{math-comp\_fourcolor}
\DefMacro{corpus-fourcolor-short-name}{fourcolor}
\DefMacro{corpus-fourcolor-url}{https://github.com/math-comp/fourcolor}
\DefMacro{corpus-fourcolor-url-pretty}{math-comp/fourcolor}
\DefMacro{corpus-fourcolor-sha}{0851d498d9a06ab81cf978494b6d41fbeb5f6631}
\DefMacro{corpus-fourcolor-sha-pretty}{0851d49}
\DefMacro{corpus-fourcolor-tag}{0851d498d9a06ab81cf978494b6d41fbeb5f6631}
\DefMacro{corpus-fourcolor-tier}{1}
\DefMacro{corpus-finmap-full-name}{math-comp\_finmap}
\DefMacro{corpus-finmap-short-name}{finmap}
\DefMacro{corpus-finmap-url}{https://github.com/math-comp/finmap}
\DefMacro{corpus-finmap-url-pretty}{math-comp/finmap}
\DefMacro{corpus-finmap-sha}{27642a81f796def99f050f2b3af41c275eac889f}
\DefMacro{corpus-finmap-sha-pretty}{27642a8}
\DefMacro{corpus-finmap-tag}{27642a81f796def99f050f2b3af41c275eac889f}
\DefMacro{corpus-finmap-tier}{1}
\DefMacro{corpus-bigenough-full-name}{math-comp\_bigenough}
\DefMacro{corpus-bigenough-short-name}{bigenough}
\DefMacro{corpus-bigenough-url}{https://github.com/math-comp/bigenough}
\DefMacro{corpus-bigenough-url-pretty}{math-comp/bigenough}
\DefMacro{corpus-bigenough-sha}{5f79a32ab69e255b9313b1e0d91577066d02bee6}
\DefMacro{corpus-bigenough-sha-pretty}{5f79a32}
\DefMacro{corpus-bigenough-tag}{5f79a32ab69e255b9313b1e0d91577066d02bee6}
\DefMacro{corpus-bigenough-tier}{2}
\DefMacro{corpus-analysis-full-name}{math-comp\_analysis}
\DefMacro{corpus-analysis-short-name}{analysis}
\DefMacro{corpus-analysis-url}{https://github.com/math-comp/analysis}
\DefMacro{corpus-analysis-url-pretty}{math-comp/analysis}
\DefMacro{corpus-analysis-sha}{9e5fe1dfa8faca954ba1c74ed301276d0e9d8339}
\DefMacro{corpus-analysis-sha-pretty}{9e5fe1d}
\DefMacro{corpus-analysis-tag}{9e5fe1dfa8faca954ba1c74ed301276d0e9d8339}
\DefMacro{corpus-analysis-tier}{2}
\DefMacro{corpus-realclosed-full-name}{math-comp\_real-closed}
\DefMacro{corpus-realclosed-short-name}{real-closed}
\DefMacro{corpus-realclosed-url}{https://github.com/math-comp/real-closed}
\DefMacro{corpus-realclosed-url-pretty}{math-comp/real-closed}
\DefMacro{corpus-realclosed-sha}{495a1faf3a13f5a5587646b6ba46061cfdeca33b}
\DefMacro{corpus-realclosed-sha-pretty}{495a1fa}
\DefMacro{corpus-realclosed-tag}{495a1faf3a13f5a5587646b6ba46061cfdeca33b}
\DefMacro{corpus-realclosed-tier}{2}
\DefMacro{corpus-robot-full-name}{affeldt-aist\_coq-robot}
\DefMacro{corpus-robot-short-name}{robot}
\DefMacro{corpus-robot-url}{https://github.com/affeldt-aist/coq-robot}
\DefMacro{corpus-robot-url-pretty}{affeldt-aist/coq-robot}
\DefMacro{corpus-robot-sha}{b341ad1dbbd7d693a1672084c97876b145792683}
\DefMacro{corpus-robot-sha-pretty}{b341ad1}
\DefMacro{corpus-robot-tag}{b341ad1dbbd7d693a1672084c97876b145792683}
\DefMacro{corpus-robot-tier}{2}
\DefMacro{corpus-multinomials-full-name}{math-comp\_multinomials}
\DefMacro{corpus-multinomials-short-name}{multinomials}
\DefMacro{corpus-multinomials-url}{https://github.com/math-comp/multinomials}
\DefMacro{corpus-multinomials-url-pretty}{math-comp/multinomials}
\DefMacro{corpus-multinomials-sha}{691d7954d3e08db172784f3e4fffe2367cd08c95}
\DefMacro{corpus-multinomials-sha-pretty}{691d795}
\DefMacro{corpus-multinomials-tag}{691d7954d3e08db172784f3e4fffe2367cd08c95}
\DefMacro{corpus-multinomials-tier}{2}
\DefMacro{corpus-ellipticcurves-full-name}{strub\_elliptic-curves-ssr}
\DefMacro{corpus-ellipticcurves-short-name}{elliptic-curves}
\DefMacro{corpus-ellipticcurves-url}{https://github.com/strub/elliptic-curves-ssr}
\DefMacro{corpus-ellipticcurves-url-pretty}{strub/elliptic-curves-ssr}
\DefMacro{corpus-ellipticcurves-sha}{631af893e591466207929714c45b5f7476d579d0}
\DefMacro{corpus-ellipticcurves-sha-pretty}{631af89}
\DefMacro{corpus-ellipticcurves-tag}{631af893e591466207929714c45b5f7476d579d0}
\DefMacro{corpus-ellipticcurves-tier}{2}
\DefMacro{corpus-toychain-full-name}{certichain\_toychain}
\DefMacro{corpus-toychain-short-name}{toychain}
\DefMacro{corpus-toychain-url}{https://github.com/certichain/toychain}
\DefMacro{corpus-toychain-url-pretty}{certichain/toychain}
\DefMacro{corpus-toychain-sha}{97bd6972f8d59eb1a56a429d59bfe10a0de6a281}
\DefMacro{corpus-toychain-sha-pretty}{97bd697}
\DefMacro{corpus-toychain-tag}{97bd6972f8d59eb1a56a429d59bfe10a0de6a281}
\DefMacro{corpus-toychain-tier}{3}
\DefMacro{corpus-bits-full-name}{coq-community\_coq-bits}
\DefMacro{corpus-bits-short-name}{bits}
\DefMacro{corpus-bits-url}{https://github.com/coq-community/coq-bits}
\DefMacro{corpus-bits-url-pretty}{coq-community/coq-bits}
\DefMacro{corpus-bits-sha}{3cd298ac7718305da6d22eaa5872b7e63651dbd0}
\DefMacro{corpus-bits-sha-pretty}{3cd298a}
\DefMacro{corpus-bits-tag}{3cd298ac7718305da6d22eaa5872b7e63651dbd0}
\DefMacro{corpus-bits-tier}{3}
\DefMacro{corpus-twosquare-full-name}{thery\_twoSquare}
\DefMacro{corpus-twosquare-short-name}{two-square}
\DefMacro{corpus-twosquare-url}{https://github.com/thery/twoSquare}
\DefMacro{corpus-twosquare-url-pretty}{thery/twoSquare}
\DefMacro{corpus-twosquare-sha}{1c09aca1080f36ec24276a2025cf87a25efdd29e}
\DefMacro{corpus-twosquare-sha-pretty}{1c09aca}
\DefMacro{corpus-twosquare-tag}{1c09aca1080f36ec24276a2025cf87a25efdd29e}
\DefMacro{corpus-twosquare-tier}{2}
\DefMacro{corpus-grobner-full-name}{thery\_grobner}
\DefMacro{corpus-grobner-short-name}{grobner}
\DefMacro{corpus-grobner-url}{https://github.com/thery/grobner}
\DefMacro{corpus-grobner-url-pretty}{thery/grobner}
\DefMacro{corpus-grobner-sha}{dfa54f9c2cd2a3913cfa4705efd9307b340197a4}
\DefMacro{corpus-grobner-sha-pretty}{dfa54f9}
\DefMacro{corpus-grobner-tag}{dfa54f9c2cd2a3913cfa4705efd9307b340197a4}
\DefMacro{corpus-grobner-tier}{2}
\DefMacro{corpus-infotheo-full-name}{palmskog\_infotheo}
\DefMacro{corpus-infotheo-short-name}{infotheo}
\DefMacro{corpus-infotheo-url}{https://github.com/palmskog/infotheo}
\DefMacro{corpus-infotheo-url-pretty}{palmskog/infotheo}
\DefMacro{corpus-infotheo-sha}{6c1724247583d084b05b092e79921465a9510e2e}
\DefMacro{corpus-infotheo-sha-pretty}{6c17242}
\DefMacro{corpus-infotheo-tag}{6c1724247583d084b05b092e79921465a9510e2e}
\DefMacro{corpus-infotheo-tier}{2}
\DefMacro{corpus-monae-full-name}{palmskog\_monae}
\DefMacro{corpus-monae-short-name}{monae}
\DefMacro{corpus-monae-url}{https://github.com/palmskog/monae}
\DefMacro{corpus-monae-url-pretty}{palmskog/monae}
\DefMacro{corpus-monae-sha}{9d0e461695e27059022271b304b75ac1bab70d3f}
\DefMacro{corpus-monae-sha-pretty}{9d0e461}
\DefMacro{corpus-monae-tag}{9d0e461695e27059022271b304b75ac1bab70d3f}
\DefMacro{corpus-monae-tier}{3}
\DefMacro{corpus-games-full-name}{gstew5\_games}
\DefMacro{corpus-games-short-name}{games}
\DefMacro{corpus-games-url}{https://github.com/gstew5/games}
\DefMacro{corpus-games-url-pretty}{gstew5/games}
\DefMacro{corpus-games-sha}{3d3bd31c7e4d1ea6f4b5a6815bca5c0a039b204f}
\DefMacro{corpus-games-sha-pretty}{3d3bd31}
\DefMacro{corpus-games-tag}{3d3bd31c7e4d1ea6f4b5a6815bca5c0a039b204f}
\DefMacro{corpus-games-tier}{3}
\DefMacro{corpus-mathcomp-num-docs}{89}
\DefMacro{corpus-mathcomp-spec-loc}{38,243}
\DefMacro{corpus-mathcomp-k-spec-loc}{38}
\DefMacro{corpus-mathcomp-proof-loc}{46,470}
\DefMacro{corpus-mathcomp-k-proof-loc}{46}
\DefMacro{corpus-mathcomp-comment-loc}{6,131}
\DefMacro{corpus-mathcomp-k-comment-loc}{6}
\DefMacro{corpus-mathcomp-total-loc}{90,844}
\DefMacro{corpus-mathcomp-k-total-loc}{90}
\DefMacro{corpus-mathcomp-code-loc}{84,713}
\DefMacro{corpus-mathcomp-k-code-loc}{84}
\DefMacro{corpus-mathcomp-AVG-spec-loc-per-doc}{429.70}
\DefMacro{corpus-mathcomp-SUM-spec-loc-per-doc}{38,243}
\DefMacro{corpus-mathcomp-MAX-spec-loc-per-doc}{4,331}
\DefMacro{corpus-mathcomp-MIN-spec-loc-per-doc}{1}
\DefMacro{corpus-mathcomp-MEDIAN-spec-loc-per-doc}{268.00}
\DefMacro{corpus-mathcomp-STDEV-spec-loc-per-doc}{567.03}
\DefMacro{corpus-mathcomp-AVG-proof-loc-per-doc}{522.13}
\DefMacro{corpus-mathcomp-SUM-proof-loc-per-doc}{46,470}
\DefMacro{corpus-mathcomp-MAX-proof-loc-per-doc}{3,084}
\DefMacro{corpus-mathcomp-MIN-proof-loc-per-doc}{0}
\DefMacro{corpus-mathcomp-MEDIAN-proof-loc-per-doc}{449.00}
\DefMacro{corpus-mathcomp-STDEV-proof-loc-per-doc}{514.98}
\DefMacro{corpus-mathcomp-AVG-comment-loc-per-doc}{68.89}
\DefMacro{corpus-mathcomp-SUM-comment-loc-per-doc}{6,131}
\DefMacro{corpus-mathcomp-MAX-comment-loc-per-doc}{588}
\DefMacro{corpus-mathcomp-MIN-comment-loc-per-doc}{0}
\DefMacro{corpus-mathcomp-MEDIAN-comment-loc-per-doc}{42.00}
\DefMacro{corpus-mathcomp-STDEV-comment-loc-per-doc}{84.63}
\DefMacro{corpus-mathcomp-AVG-total-loc-per-doc}{1,020.72}
\DefMacro{corpus-mathcomp-SUM-total-loc-per-doc}{90,844}
\DefMacro{corpus-mathcomp-MAX-total-loc-per-doc}{5,619}
\DefMacro{corpus-mathcomp-MIN-total-loc-per-doc}{1}
\DefMacro{corpus-mathcomp-MEDIAN-total-loc-per-doc}{729.00}
\DefMacro{corpus-mathcomp-STDEV-total-loc-per-doc}{1,008.01}
\DefMacro{corpus-mathcomp-AVG-code-loc-per-doc}{951.83}
\DefMacro{corpus-mathcomp-SUM-code-loc-per-doc}{84,713}
\DefMacro{corpus-mathcomp-MAX-code-loc-per-doc}{5,031}
\DefMacro{corpus-mathcomp-MIN-code-loc-per-doc}{1}
\DefMacro{corpus-mathcomp-MEDIAN-code-loc-per-doc}{675.00}
\DefMacro{corpus-mathcomp-STDEV-code-loc-per-doc}{936.58}
\DefMacro{corpus-fcslpcm-num-docs}{12}
\DefMacro{corpus-fcslpcm-spec-loc}{2,937}
\DefMacro{corpus-fcslpcm-k-spec-loc}{2}
\DefMacro{corpus-fcslpcm-proof-loc}{2,852}
\DefMacro{corpus-fcslpcm-k-proof-loc}{2}
\DefMacro{corpus-fcslpcm-comment-loc}{466}
\DefMacro{corpus-fcslpcm-k-comment-loc}{0}
\DefMacro{corpus-fcslpcm-total-loc}{6,255}
\DefMacro{corpus-fcslpcm-k-total-loc}{6}
\DefMacro{corpus-fcslpcm-code-loc}{5,789}
\DefMacro{corpus-fcslpcm-k-code-loc}{5}
\DefMacro{corpus-fcslpcm-AVG-spec-loc-per-doc}{244.75}
\DefMacro{corpus-fcslpcm-SUM-spec-loc-per-doc}{2,937}
\DefMacro{corpus-fcslpcm-MAX-spec-loc-per-doc}{722}
\DefMacro{corpus-fcslpcm-MIN-spec-loc-per-doc}{38}
\DefMacro{corpus-fcslpcm-MEDIAN-spec-loc-per-doc}{224.00}
\DefMacro{corpus-fcslpcm-STDEV-spec-loc-per-doc}{172.58}
\DefMacro{corpus-fcslpcm-AVG-proof-loc-per-doc}{237.67}
\DefMacro{corpus-fcslpcm-SUM-proof-loc-per-doc}{2,852}
\DefMacro{corpus-fcslpcm-MAX-proof-loc-per-doc}{981}
\DefMacro{corpus-fcslpcm-MIN-proof-loc-per-doc}{25}
\DefMacro{corpus-fcslpcm-MEDIAN-proof-loc-per-doc}{119.50}
\DefMacro{corpus-fcslpcm-STDEV-proof-loc-per-doc}{287.34}
\DefMacro{corpus-fcslpcm-AVG-comment-loc-per-doc}{38.83}
\DefMacro{corpus-fcslpcm-SUM-comment-loc-per-doc}{466}
\DefMacro{corpus-fcslpcm-MAX-comment-loc-per-doc}{126}
\DefMacro{corpus-fcslpcm-MIN-comment-loc-per-doc}{7}
\DefMacro{corpus-fcslpcm-MEDIAN-comment-loc-per-doc}{19.50}
\DefMacro{corpus-fcslpcm-STDEV-comment-loc-per-doc}{39.02}
\DefMacro{corpus-fcslpcm-AVG-total-loc-per-doc}{521.25}
\DefMacro{corpus-fcslpcm-SUM-total-loc-per-doc}{6,255}
\DefMacro{corpus-fcslpcm-MAX-total-loc-per-doc}{1,821}
\DefMacro{corpus-fcslpcm-MIN-total-loc-per-doc}{77}
\DefMacro{corpus-fcslpcm-MEDIAN-total-loc-per-doc}{427.00}
\DefMacro{corpus-fcslpcm-STDEV-total-loc-per-doc}{470.64}
\DefMacro{corpus-fcslpcm-AVG-code-loc-per-doc}{482.42}
\DefMacro{corpus-fcslpcm-SUM-code-loc-per-doc}{5,789}
\DefMacro{corpus-fcslpcm-MAX-code-loc-per-doc}{1,703}
\DefMacro{corpus-fcslpcm-MIN-code-loc-per-doc}{63}
\DefMacro{corpus-fcslpcm-MEDIAN-code-loc-per-doc}{383.00}
\DefMacro{corpus-fcslpcm-STDEV-code-loc-per-doc}{447.18}
\DefMacro{corpus-disel-num-docs}{20}
\DefMacro{corpus-disel-spec-loc}{2,575}
\DefMacro{corpus-disel-k-spec-loc}{2}
\DefMacro{corpus-disel-proof-loc}{1,898}
\DefMacro{corpus-disel-k-proof-loc}{1}
\DefMacro{corpus-disel-comment-loc}{505}
\DefMacro{corpus-disel-k-comment-loc}{0}
\DefMacro{corpus-disel-total-loc}{4,978}
\DefMacro{corpus-disel-k-total-loc}{4}
\DefMacro{corpus-disel-code-loc}{4,473}
\DefMacro{corpus-disel-k-code-loc}{4}
\DefMacro{corpus-disel-AVG-spec-loc-per-doc}{128.75}
\DefMacro{corpus-disel-SUM-spec-loc-per-doc}{2,575}
\DefMacro{corpus-disel-MAX-spec-loc-per-doc}{610}
\DefMacro{corpus-disel-MIN-spec-loc-per-doc}{17}
\DefMacro{corpus-disel-MEDIAN-spec-loc-per-doc}{89.00}
\DefMacro{corpus-disel-STDEV-spec-loc-per-doc}{125.21}
\DefMacro{corpus-disel-AVG-proof-loc-per-doc}{94.90}
\DefMacro{corpus-disel-SUM-proof-loc-per-doc}{1,898}
\DefMacro{corpus-disel-MAX-proof-loc-per-doc}{315}
\DefMacro{corpus-disel-MIN-proof-loc-per-doc}{0}
\DefMacro{corpus-disel-MEDIAN-proof-loc-per-doc}{60.00}
\DefMacro{corpus-disel-STDEV-proof-loc-per-doc}{89.67}
\DefMacro{corpus-disel-AVG-comment-loc-per-doc}{25.25}
\DefMacro{corpus-disel-SUM-comment-loc-per-doc}{505}
\DefMacro{corpus-disel-MAX-comment-loc-per-doc}{84}
\DefMacro{corpus-disel-MIN-comment-loc-per-doc}{0}
\DefMacro{corpus-disel-MEDIAN-comment-loc-per-doc}{23.00}
\DefMacro{corpus-disel-STDEV-comment-loc-per-doc}{22.87}
\DefMacro{corpus-disel-AVG-total-loc-per-doc}{248.90}
\DefMacro{corpus-disel-SUM-total-loc-per-doc}{4,978}
\DefMacro{corpus-disel-MAX-total-loc-per-doc}{939}
\DefMacro{corpus-disel-MIN-total-loc-per-doc}{17}
\DefMacro{corpus-disel-MEDIAN-total-loc-per-doc}{204.50}
\DefMacro{corpus-disel-STDEV-total-loc-per-doc}{205.41}
\DefMacro{corpus-disel-AVG-code-loc-per-doc}{223.65}
\DefMacro{corpus-disel-SUM-code-loc-per-doc}{4,473}
\DefMacro{corpus-disel-MAX-code-loc-per-doc}{855}
\DefMacro{corpus-disel-MIN-code-loc-per-doc}{17}
\DefMacro{corpus-disel-MEDIAN-code-loc-per-doc}{170.50}
\DefMacro{corpus-disel-STDEV-code-loc-per-doc}{190.65}
\DefMacro{corpus-reglang-num-docs}{12}
\DefMacro{corpus-reglang-spec-loc}{1,299}
\DefMacro{corpus-reglang-k-spec-loc}{1}
\DefMacro{corpus-reglang-proof-loc}{1,734}
\DefMacro{corpus-reglang-k-proof-loc}{1}
\DefMacro{corpus-reglang-comment-loc}{250}
\DefMacro{corpus-reglang-k-comment-loc}{0}
\DefMacro{corpus-reglang-total-loc}{3,283}
\DefMacro{corpus-reglang-k-total-loc}{3}
\DefMacro{corpus-reglang-code-loc}{3,033}
\DefMacro{corpus-reglang-k-code-loc}{3}
\DefMacro{corpus-reglang-AVG-spec-loc-per-doc}{108.25}
\DefMacro{corpus-reglang-SUM-spec-loc-per-doc}{1,299}
\DefMacro{corpus-reglang-MAX-spec-loc-per-doc}{273}
\DefMacro{corpus-reglang-MIN-spec-loc-per-doc}{20}
\DefMacro{corpus-reglang-MEDIAN-spec-loc-per-doc}{90.00}
\DefMacro{corpus-reglang-STDEV-spec-loc-per-doc}{67.77}
\DefMacro{corpus-reglang-AVG-proof-loc-per-doc}{144.50}
\DefMacro{corpus-reglang-SUM-proof-loc-per-doc}{1,734}
\DefMacro{corpus-reglang-MAX-proof-loc-per-doc}{506}
\DefMacro{corpus-reglang-MIN-proof-loc-per-doc}{13}
\DefMacro{corpus-reglang-MEDIAN-proof-loc-per-doc}{101.00}
\DefMacro{corpus-reglang-STDEV-proof-loc-per-doc}{126.67}
\DefMacro{corpus-reglang-AVG-comment-loc-per-doc}{20.83}
\DefMacro{corpus-reglang-SUM-comment-loc-per-doc}{250}
\DefMacro{corpus-reglang-MAX-comment-loc-per-doc}{38}
\DefMacro{corpus-reglang-MIN-comment-loc-per-doc}{2}
\DefMacro{corpus-reglang-MEDIAN-comment-loc-per-doc}{19.50}
\DefMacro{corpus-reglang-STDEV-comment-loc-per-doc}{10.50}
\DefMacro{corpus-reglang-AVG-total-loc-per-doc}{273.58}
\DefMacro{corpus-reglang-SUM-total-loc-per-doc}{3,283}
\DefMacro{corpus-reglang-MAX-total-loc-per-doc}{816}
\DefMacro{corpus-reglang-MIN-total-loc-per-doc}{39}
\DefMacro{corpus-reglang-MEDIAN-total-loc-per-doc}{211.50}
\DefMacro{corpus-reglang-STDEV-total-loc-per-doc}{199.27}
\DefMacro{corpus-reglang-AVG-code-loc-per-doc}{252.75}
\DefMacro{corpus-reglang-SUM-code-loc-per-doc}{3,033}
\DefMacro{corpus-reglang-MAX-code-loc-per-doc}{779}
\DefMacro{corpus-reglang-MIN-code-loc-per-doc}{33}
\DefMacro{corpus-reglang-MEDIAN-code-loc-per-doc}{192.50}
\DefMacro{corpus-reglang-STDEV-code-loc-per-doc}{191.22}
\DefMacro{corpus-compdecpdl-num-docs}{16}
\DefMacro{corpus-compdecpdl-spec-loc}{2,305}
\DefMacro{corpus-compdecpdl-k-spec-loc}{2}
\DefMacro{corpus-compdecpdl-proof-loc}{2,114}
\DefMacro{corpus-compdecpdl-k-proof-loc}{2}
\DefMacro{corpus-compdecpdl-comment-loc}{394}
\DefMacro{corpus-compdecpdl-k-comment-loc}{0}
\DefMacro{corpus-compdecpdl-total-loc}{4,813}
\DefMacro{corpus-compdecpdl-k-total-loc}{4}
\DefMacro{corpus-compdecpdl-code-loc}{4,419}
\DefMacro{corpus-compdecpdl-k-code-loc}{4}
\DefMacro{corpus-compdecpdl-AVG-spec-loc-per-doc}{144.06}
\DefMacro{corpus-compdecpdl-SUM-spec-loc-per-doc}{2,305}
\DefMacro{corpus-compdecpdl-MAX-spec-loc-per-doc}{464}
\DefMacro{corpus-compdecpdl-MIN-spec-loc-per-doc}{18}
\DefMacro{corpus-compdecpdl-MEDIAN-spec-loc-per-doc}{78.50}
\DefMacro{corpus-compdecpdl-STDEV-spec-loc-per-doc}{145.99}
\DefMacro{corpus-compdecpdl-AVG-proof-loc-per-doc}{132.12}
\DefMacro{corpus-compdecpdl-SUM-proof-loc-per-doc}{2,114}
\DefMacro{corpus-compdecpdl-MAX-proof-loc-per-doc}{415}
\DefMacro{corpus-compdecpdl-MIN-proof-loc-per-doc}{0}
\DefMacro{corpus-compdecpdl-MEDIAN-proof-loc-per-doc}{82.50}
\DefMacro{corpus-compdecpdl-STDEV-proof-loc-per-doc}{128.70}
\DefMacro{corpus-compdecpdl-AVG-comment-loc-per-doc}{24.62}
\DefMacro{corpus-compdecpdl-SUM-comment-loc-per-doc}{394}
\DefMacro{corpus-compdecpdl-MAX-comment-loc-per-doc}{89}
\DefMacro{corpus-compdecpdl-MIN-comment-loc-per-doc}{1}
\DefMacro{corpus-compdecpdl-MEDIAN-comment-loc-per-doc}{12.00}
\DefMacro{corpus-compdecpdl-STDEV-comment-loc-per-doc}{25.94}
\DefMacro{corpus-compdecpdl-AVG-total-loc-per-doc}{300.81}
\DefMacro{corpus-compdecpdl-SUM-total-loc-per-doc}{4,813}
\DefMacro{corpus-compdecpdl-MAX-total-loc-per-doc}{891}
\DefMacro{corpus-compdecpdl-MIN-total-loc-per-doc}{22}
\DefMacro{corpus-compdecpdl-MEDIAN-total-loc-per-doc}{211.50}
\DefMacro{corpus-compdecpdl-STDEV-total-loc-per-doc}{283.14}
\DefMacro{corpus-compdecpdl-AVG-code-loc-per-doc}{276.19}
\DefMacro{corpus-compdecpdl-SUM-code-loc-per-doc}{4,419}
\DefMacro{corpus-compdecpdl-MAX-code-loc-per-doc}{802}
\DefMacro{corpus-compdecpdl-MIN-code-loc-per-doc}{19}
\DefMacro{corpus-compdecpdl-MEDIAN-code-loc-per-doc}{199.50}
\DefMacro{corpus-compdecpdl-STDEV-code-loc-per-doc}{259.82}
\DefMacro{corpus-oddorder-num-docs}{34}
\DefMacro{corpus-oddorder-spec-loc}{11,882}
\DefMacro{corpus-oddorder-k-spec-loc}{11}
\DefMacro{corpus-oddorder-proof-loc}{24,243}
\DefMacro{corpus-oddorder-k-proof-loc}{24}
\DefMacro{corpus-oddorder-comment-loc}{2,313}
\DefMacro{corpus-oddorder-k-comment-loc}{2}
\DefMacro{corpus-oddorder-total-loc}{38,438}
\DefMacro{corpus-oddorder-k-total-loc}{38}
\DefMacro{corpus-oddorder-code-loc}{36,125}
\DefMacro{corpus-oddorder-k-code-loc}{36}
\DefMacro{corpus-oddorder-AVG-spec-loc-per-doc}{349.47}
\DefMacro{corpus-oddorder-SUM-spec-loc-per-doc}{11,882}
\DefMacro{corpus-oddorder-MAX-spec-loc-per-doc}{889}
\DefMacro{corpus-oddorder-MIN-spec-loc-per-doc}{30}
\DefMacro{corpus-oddorder-MEDIAN-spec-loc-per-doc}{285.50}
\DefMacro{corpus-oddorder-STDEV-spec-loc-per-doc}{247.22}
\DefMacro{corpus-oddorder-AVG-proof-loc-per-doc}{713.03}
\DefMacro{corpus-oddorder-SUM-proof-loc-per-doc}{24,243}
\DefMacro{corpus-oddorder-MAX-proof-loc-per-doc}{1,829}
\DefMacro{corpus-oddorder-MIN-proof-loc-per-doc}{104}
\DefMacro{corpus-oddorder-MEDIAN-proof-loc-per-doc}{600.50}
\DefMacro{corpus-oddorder-STDEV-proof-loc-per-doc}{433.11}
\DefMacro{corpus-oddorder-AVG-comment-loc-per-doc}{68.03}
\DefMacro{corpus-oddorder-SUM-comment-loc-per-doc}{2,313}
\DefMacro{corpus-oddorder-MAX-comment-loc-per-doc}{191}
\DefMacro{corpus-oddorder-MIN-comment-loc-per-doc}{2}
\DefMacro{corpus-oddorder-MEDIAN-comment-loc-per-doc}{55.50}
\DefMacro{corpus-oddorder-STDEV-comment-loc-per-doc}{46.93}
\DefMacro{corpus-oddorder-AVG-total-loc-per-doc}{1,130.53}
\DefMacro{corpus-oddorder-SUM-total-loc-per-doc}{38,438}
\DefMacro{corpus-oddorder-MAX-total-loc-per-doc}{2,662}
\DefMacro{corpus-oddorder-MIN-total-loc-per-doc}{176}
\DefMacro{corpus-oddorder-MEDIAN-total-loc-per-doc}{1,124.00}
\DefMacro{corpus-oddorder-STDEV-total-loc-per-doc}{600.21}
\DefMacro{corpus-oddorder-AVG-code-loc-per-doc}{1,062.50}
\DefMacro{corpus-oddorder-SUM-code-loc-per-doc}{36,125}
\DefMacro{corpus-oddorder-MAX-code-loc-per-doc}{2,526}
\DefMacro{corpus-oddorder-MIN-code-loc-per-doc}{164}
\DefMacro{corpus-oddorder-MEDIAN-code-loc-per-doc}{1,060.50}
\DefMacro{corpus-oddorder-STDEV-code-loc-per-doc}{569.58}
\DefMacro{corpus-fourcolor-num-docs}{60}
\DefMacro{corpus-fourcolor-spec-loc}{9,175}
\DefMacro{corpus-fourcolor-k-spec-loc}{9}
\DefMacro{corpus-fourcolor-proof-loc}{27,963}
\DefMacro{corpus-fourcolor-k-proof-loc}{27}
\DefMacro{corpus-fourcolor-comment-loc}{2,816}
\DefMacro{corpus-fourcolor-k-comment-loc}{2}
\DefMacro{corpus-fourcolor-total-loc}{39,954}
\DefMacro{corpus-fourcolor-k-total-loc}{39}
\DefMacro{corpus-fourcolor-code-loc}{37,138}
\DefMacro{corpus-fourcolor-k-code-loc}{37}
\DefMacro{corpus-fourcolor-AVG-spec-loc-per-doc}{152.92}
\DefMacro{corpus-fourcolor-SUM-spec-loc-per-doc}{9,175}
\DefMacro{corpus-fourcolor-MAX-spec-loc-per-doc}{1,486}
\DefMacro{corpus-fourcolor-MIN-spec-loc-per-doc}{6}
\DefMacro{corpus-fourcolor-MEDIAN-spec-loc-per-doc}{115.00}
\DefMacro{corpus-fourcolor-STDEV-spec-loc-per-doc}{214.81}
\DefMacro{corpus-fourcolor-AVG-proof-loc-per-doc}{466.05}
\DefMacro{corpus-fourcolor-SUM-proof-loc-per-doc}{27,963}
\DefMacro{corpus-fourcolor-MAX-proof-loc-per-doc}{5,551}
\DefMacro{corpus-fourcolor-MIN-proof-loc-per-doc}{0}
\DefMacro{corpus-fourcolor-MEDIAN-proof-loc-per-doc}{232.00}
\DefMacro{corpus-fourcolor-STDEV-proof-loc-per-doc}{1,006.70}
\DefMacro{corpus-fourcolor-AVG-comment-loc-per-doc}{46.93}
\DefMacro{corpus-fourcolor-SUM-comment-loc-per-doc}{2,816}
\DefMacro{corpus-fourcolor-MAX-comment-loc-per-doc}{178}
\DefMacro{corpus-fourcolor-MIN-comment-loc-per-doc}{2}
\DefMacro{corpus-fourcolor-MEDIAN-comment-loc-per-doc}{41.50}
\DefMacro{corpus-fourcolor-STDEV-comment-loc-per-doc}{39.47}
\DefMacro{corpus-fourcolor-AVG-total-loc-per-doc}{665.90}
\DefMacro{corpus-fourcolor-SUM-total-loc-per-doc}{39,954}
\DefMacro{corpus-fourcolor-MAX-total-loc-per-doc}{5,563}
\DefMacro{corpus-fourcolor-MIN-total-loc-per-doc}{17}
\DefMacro{corpus-fourcolor-MEDIAN-total-loc-per-doc}{398.50}
\DefMacro{corpus-fourcolor-STDEV-total-loc-per-doc}{994.17}
\DefMacro{corpus-fourcolor-AVG-code-loc-per-doc}{618.97}
\DefMacro{corpus-fourcolor-SUM-code-loc-per-doc}{37,138}
\DefMacro{corpus-fourcolor-MAX-code-loc-per-doc}{5,560}
\DefMacro{corpus-fourcolor-MIN-code-loc-per-doc}{15}
\DefMacro{corpus-fourcolor-MEDIAN-code-loc-per-doc}{343.00}
\DefMacro{corpus-fourcolor-STDEV-code-loc-per-doc}{999.58}
\DefMacro{corpus-finmap-num-docs}{4}
\DefMacro{corpus-finmap-spec-loc}{4,260}
\DefMacro{corpus-finmap-k-spec-loc}{4}
\DefMacro{corpus-finmap-proof-loc}{2,191}
\DefMacro{corpus-finmap-k-proof-loc}{2}
\DefMacro{corpus-finmap-comment-loc}{238}
\DefMacro{corpus-finmap-k-comment-loc}{0}
\DefMacro{corpus-finmap-total-loc}{6,689}
\DefMacro{corpus-finmap-k-total-loc}{6}
\DefMacro{corpus-finmap-code-loc}{6,451}
\DefMacro{corpus-finmap-k-code-loc}{6}
\DefMacro{corpus-finmap-AVG-spec-loc-per-doc}{1,065.00}
\DefMacro{corpus-finmap-SUM-spec-loc-per-doc}{4,260}
\DefMacro{corpus-finmap-MAX-spec-loc-per-doc}{1,831}
\DefMacro{corpus-finmap-MIN-spec-loc-per-doc}{338}
\DefMacro{corpus-finmap-MEDIAN-spec-loc-per-doc}{1,045.50}
\DefMacro{corpus-finmap-STDEV-spec-loc-per-doc}{651.57}
\DefMacro{corpus-finmap-AVG-proof-loc-per-doc}{547.75}
\DefMacro{corpus-finmap-SUM-proof-loc-per-doc}{2,191}
\DefMacro{corpus-finmap-MAX-proof-loc-per-doc}{1,184}
\DefMacro{corpus-finmap-MIN-proof-loc-per-doc}{264}
\DefMacro{corpus-finmap-MEDIAN-proof-loc-per-doc}{371.50}
\DefMacro{corpus-finmap-STDEV-proof-loc-per-doc}{370.29}
\DefMacro{corpus-finmap-AVG-comment-loc-per-doc}{59.50}
\DefMacro{corpus-finmap-SUM-comment-loc-per-doc}{238}
\DefMacro{corpus-finmap-MAX-comment-loc-per-doc}{136}
\DefMacro{corpus-finmap-MIN-comment-loc-per-doc}{18}
\DefMacro{corpus-finmap-MEDIAN-comment-loc-per-doc}{42.00}
\DefMacro{corpus-finmap-STDEV-comment-loc-per-doc}{45.90}
\DefMacro{corpus-finmap-AVG-total-loc-per-doc}{1,672.25}
\DefMacro{corpus-finmap-SUM-total-loc-per-doc}{6,689}
\DefMacro{corpus-finmap-MAX-total-loc-per-doc}{2,905}
\DefMacro{corpus-finmap-MIN-total-loc-per-doc}{718}
\DefMacro{corpus-finmap-MEDIAN-total-loc-per-doc}{1,533.00}
\DefMacro{corpus-finmap-STDEV-total-loc-per-doc}{945.93}
\DefMacro{corpus-finmap-AVG-code-loc-per-doc}{1,612.75}
\DefMacro{corpus-finmap-SUM-code-loc-per-doc}{6,451}
\DefMacro{corpus-finmap-MAX-code-loc-per-doc}{2,769}
\DefMacro{corpus-finmap-MIN-code-loc-per-doc}{687}
\DefMacro{corpus-finmap-MEDIAN-code-loc-per-doc}{1,497.50}
\DefMacro{corpus-finmap-STDEV-code-loc-per-doc}{905.40}
\DefMacro{corpus-bigenough-num-docs}{1}
\DefMacro{corpus-bigenough-spec-loc}{70}
\DefMacro{corpus-bigenough-k-spec-loc}{0}
\DefMacro{corpus-bigenough-proof-loc}{8}
\DefMacro{corpus-bigenough-k-proof-loc}{0}
\DefMacro{corpus-bigenough-comment-loc}{16}
\DefMacro{corpus-bigenough-k-comment-loc}{0}
\DefMacro{corpus-bigenough-total-loc}{94}
\DefMacro{corpus-bigenough-k-total-loc}{0}
\DefMacro{corpus-bigenough-code-loc}{78}
\DefMacro{corpus-bigenough-k-code-loc}{0}
\DefMacro{corpus-bigenough-AVG-spec-loc-per-doc}{70.00}
\DefMacro{corpus-bigenough-SUM-spec-loc-per-doc}{70}
\DefMacro{corpus-bigenough-MAX-spec-loc-per-doc}{70}
\DefMacro{corpus-bigenough-MIN-spec-loc-per-doc}{70}
\DefMacro{corpus-bigenough-MEDIAN-spec-loc-per-doc}{70.00}
\DefMacro{corpus-bigenough-STDEV-spec-loc-per-doc}{0.00}
\DefMacro{corpus-bigenough-AVG-proof-loc-per-doc}{8.00}
\DefMacro{corpus-bigenough-SUM-proof-loc-per-doc}{8}
\DefMacro{corpus-bigenough-MAX-proof-loc-per-doc}{8}
\DefMacro{corpus-bigenough-MIN-proof-loc-per-doc}{8}
\DefMacro{corpus-bigenough-MEDIAN-proof-loc-per-doc}{8.00}
\DefMacro{corpus-bigenough-STDEV-proof-loc-per-doc}{0.00}
\DefMacro{corpus-bigenough-AVG-comment-loc-per-doc}{16.00}
\DefMacro{corpus-bigenough-SUM-comment-loc-per-doc}{16}
\DefMacro{corpus-bigenough-MAX-comment-loc-per-doc}{16}
\DefMacro{corpus-bigenough-MIN-comment-loc-per-doc}{16}
\DefMacro{corpus-bigenough-MEDIAN-comment-loc-per-doc}{16.00}
\DefMacro{corpus-bigenough-STDEV-comment-loc-per-doc}{0.00}
\DefMacro{corpus-bigenough-AVG-total-loc-per-doc}{94.00}
\DefMacro{corpus-bigenough-SUM-total-loc-per-doc}{94}
\DefMacro{corpus-bigenough-MAX-total-loc-per-doc}{94}
\DefMacro{corpus-bigenough-MIN-total-loc-per-doc}{94}
\DefMacro{corpus-bigenough-MEDIAN-total-loc-per-doc}{94.00}
\DefMacro{corpus-bigenough-STDEV-total-loc-per-doc}{0.00}
\DefMacro{corpus-bigenough-AVG-code-loc-per-doc}{78.00}
\DefMacro{corpus-bigenough-SUM-code-loc-per-doc}{78}
\DefMacro{corpus-bigenough-MAX-code-loc-per-doc}{78}
\DefMacro{corpus-bigenough-MIN-code-loc-per-doc}{78}
\DefMacro{corpus-bigenough-MEDIAN-code-loc-per-doc}{78.00}
\DefMacro{corpus-bigenough-STDEV-code-loc-per-doc}{0.00}
\DefMacro{corpus-analysis-num-docs}{17}
\DefMacro{corpus-analysis-spec-loc}{5,553}
\DefMacro{corpus-analysis-k-spec-loc}{5}
\DefMacro{corpus-analysis-proof-loc}{6,186}
\DefMacro{corpus-analysis-k-proof-loc}{6}
\DefMacro{corpus-analysis-comment-loc}{1,352}
\DefMacro{corpus-analysis-k-comment-loc}{1}
\DefMacro{corpus-analysis-total-loc}{13,091}
\DefMacro{corpus-analysis-k-total-loc}{13}
\DefMacro{corpus-analysis-code-loc}{11,739}
\DefMacro{corpus-analysis-k-code-loc}{11}
\DefMacro{corpus-analysis-AVG-spec-loc-per-doc}{326.65}
\DefMacro{corpus-analysis-SUM-spec-loc-per-doc}{5,553}
\DefMacro{corpus-analysis-MAX-spec-loc-per-doc}{1,196}
\DefMacro{corpus-analysis-MIN-spec-loc-per-doc}{29}
\DefMacro{corpus-analysis-MEDIAN-spec-loc-per-doc}{256.00}
\DefMacro{corpus-analysis-STDEV-spec-loc-per-doc}{291.79}
\DefMacro{corpus-analysis-AVG-proof-loc-per-doc}{363.88}
\DefMacro{corpus-analysis-SUM-proof-loc-per-doc}{6,186}
\DefMacro{corpus-analysis-MAX-proof-loc-per-doc}{986}
\DefMacro{corpus-analysis-MIN-proof-loc-per-doc}{5}
\DefMacro{corpus-analysis-MEDIAN-proof-loc-per-doc}{282.00}
\DefMacro{corpus-analysis-STDEV-proof-loc-per-doc}{307.69}
\DefMacro{corpus-analysis-AVG-comment-loc-per-doc}{79.53}
\DefMacro{corpus-analysis-SUM-comment-loc-per-doc}{1,352}
\DefMacro{corpus-analysis-MAX-comment-loc-per-doc}{311}
\DefMacro{corpus-analysis-MIN-comment-loc-per-doc}{3}
\DefMacro{corpus-analysis-MEDIAN-comment-loc-per-doc}{36.00}
\DefMacro{corpus-analysis-STDEV-comment-loc-per-doc}{92.70}
\DefMacro{corpus-analysis-AVG-total-loc-per-doc}{770.06}
\DefMacro{corpus-analysis-SUM-total-loc-per-doc}{13,091}
\DefMacro{corpus-analysis-MAX-total-loc-per-doc}{2,493}
\DefMacro{corpus-analysis-MIN-total-loc-per-doc}{37}
\DefMacro{corpus-analysis-MEDIAN-total-loc-per-doc}{503.00}
\DefMacro{corpus-analysis-STDEV-total-loc-per-doc}{632.56}
\DefMacro{corpus-analysis-AVG-code-loc-per-doc}{690.53}
\DefMacro{corpus-analysis-SUM-code-loc-per-doc}{11,739}
\DefMacro{corpus-analysis-MAX-code-loc-per-doc}{2,182}
\DefMacro{corpus-analysis-MIN-code-loc-per-doc}{34}
\DefMacro{corpus-analysis-MEDIAN-code-loc-per-doc}{451.00}
\DefMacro{corpus-analysis-STDEV-code-loc-per-doc}{560.05}
\DefMacro{corpus-realclosed-num-docs}{10}
\DefMacro{corpus-realclosed-spec-loc}{4,348}
\DefMacro{corpus-realclosed-k-spec-loc}{4}
\DefMacro{corpus-realclosed-proof-loc}{4,577}
\DefMacro{corpus-realclosed-k-proof-loc}{4}
\DefMacro{corpus-realclosed-comment-loc}{356}
\DefMacro{corpus-realclosed-k-comment-loc}{0}
\DefMacro{corpus-realclosed-total-loc}{9,281}
\DefMacro{corpus-realclosed-k-total-loc}{9}
\DefMacro{corpus-realclosed-code-loc}{8,925}
\DefMacro{corpus-realclosed-k-code-loc}{8}
\DefMacro{corpus-realclosed-AVG-spec-loc-per-doc}{434.80}
\DefMacro{corpus-realclosed-SUM-spec-loc-per-doc}{4,348}
\DefMacro{corpus-realclosed-MAX-spec-loc-per-doc}{1,148}
\DefMacro{corpus-realclosed-MIN-spec-loc-per-doc}{8}
\DefMacro{corpus-realclosed-MEDIAN-spec-loc-per-doc}{382.00}
\DefMacro{corpus-realclosed-STDEV-spec-loc-per-doc}{355.87}
\DefMacro{corpus-realclosed-AVG-proof-loc-per-doc}{457.70}
\DefMacro{corpus-realclosed-SUM-proof-loc-per-doc}{4,577}
\DefMacro{corpus-realclosed-MAX-proof-loc-per-doc}{1,073}
\DefMacro{corpus-realclosed-MIN-proof-loc-per-doc}{0}
\DefMacro{corpus-realclosed-MEDIAN-proof-loc-per-doc}{420.50}
\DefMacro{corpus-realclosed-STDEV-proof-loc-per-doc}{421.37}
\DefMacro{corpus-realclosed-AVG-comment-loc-per-doc}{35.60}
\DefMacro{corpus-realclosed-SUM-comment-loc-per-doc}{356}
\DefMacro{corpus-realclosed-MAX-comment-loc-per-doc}{111}
\DefMacro{corpus-realclosed-MIN-comment-loc-per-doc}{0}
\DefMacro{corpus-realclosed-MEDIAN-comment-loc-per-doc}{34.50}
\DefMacro{corpus-realclosed-STDEV-comment-loc-per-doc}{35.96}
\DefMacro{corpus-realclosed-AVG-total-loc-per-doc}{928.10}
\DefMacro{corpus-realclosed-SUM-total-loc-per-doc}{9,281}
\DefMacro{corpus-realclosed-MAX-total-loc-per-doc}{1,607}
\DefMacro{corpus-realclosed-MIN-total-loc-per-doc}{8}
\DefMacro{corpus-realclosed-MEDIAN-total-loc-per-doc}{1,148.50}
\DefMacro{corpus-realclosed-STDEV-total-loc-per-doc}{530.94}
\DefMacro{corpus-realclosed-AVG-code-loc-per-doc}{892.50}
\DefMacro{corpus-realclosed-SUM-code-loc-per-doc}{8,925}
\DefMacro{corpus-realclosed-MAX-code-loc-per-doc}{1,496}
\DefMacro{corpus-realclosed-MIN-code-loc-per-doc}{8}
\DefMacro{corpus-realclosed-MEDIAN-code-loc-per-doc}{1,091.00}
\DefMacro{corpus-realclosed-STDEV-code-loc-per-doc}{509.72}
\DefMacro{corpus-robot-num-docs}{13}
\DefMacro{corpus-robot-spec-loc}{3,881}
\DefMacro{corpus-robot-k-spec-loc}{3}
\DefMacro{corpus-robot-proof-loc}{7,249}
\DefMacro{corpus-robot-k-proof-loc}{7}
\DefMacro{corpus-robot-comment-loc}{648}
\DefMacro{corpus-robot-k-comment-loc}{0}
\DefMacro{corpus-robot-total-loc}{11,778}
\DefMacro{corpus-robot-k-total-loc}{11}
\DefMacro{corpus-robot-code-loc}{11,130}
\DefMacro{corpus-robot-k-code-loc}{11}
\DefMacro{corpus-robot-AVG-spec-loc-per-doc}{298.54}
\DefMacro{corpus-robot-SUM-spec-loc-per-doc}{3,881}
\DefMacro{corpus-robot-MAX-spec-loc-per-doc}{544}
\DefMacro{corpus-robot-MIN-spec-loc-per-doc}{72}
\DefMacro{corpus-robot-MEDIAN-spec-loc-per-doc}{314.00}
\DefMacro{corpus-robot-STDEV-spec-loc-per-doc}{130.85}
\DefMacro{corpus-robot-AVG-proof-loc-per-doc}{557.62}
\DefMacro{corpus-robot-SUM-proof-loc-per-doc}{7,249}
\DefMacro{corpus-robot-MAX-proof-loc-per-doc}{1,466}
\DefMacro{corpus-robot-MIN-proof-loc-per-doc}{53}
\DefMacro{corpus-robot-MEDIAN-proof-loc-per-doc}{532.00}
\DefMacro{corpus-robot-STDEV-proof-loc-per-doc}{351.84}
\DefMacro{corpus-robot-AVG-comment-loc-per-doc}{49.85}
\DefMacro{corpus-robot-SUM-comment-loc-per-doc}{648}
\DefMacro{corpus-robot-MAX-comment-loc-per-doc}{110}
\DefMacro{corpus-robot-MIN-comment-loc-per-doc}{6}
\DefMacro{corpus-robot-MEDIAN-comment-loc-per-doc}{52.00}
\DefMacro{corpus-robot-STDEV-comment-loc-per-doc}{31.56}
\DefMacro{corpus-robot-AVG-total-loc-per-doc}{906.00}
\DefMacro{corpus-robot-SUM-total-loc-per-doc}{11,778}
\DefMacro{corpus-robot-MAX-total-loc-per-doc}{1,912}
\DefMacro{corpus-robot-MIN-total-loc-per-doc}{131}
\DefMacro{corpus-robot-MEDIAN-total-loc-per-doc}{856.00}
\DefMacro{corpus-robot-STDEV-total-loc-per-doc}{474.47}
\DefMacro{corpus-robot-AVG-code-loc-per-doc}{856.15}
\DefMacro{corpus-robot-SUM-code-loc-per-doc}{11,130}
\DefMacro{corpus-robot-MAX-code-loc-per-doc}{1,858}
\DefMacro{corpus-robot-MIN-code-loc-per-doc}{125}
\DefMacro{corpus-robot-MEDIAN-code-loc-per-doc}{819.00}
\DefMacro{corpus-robot-STDEV-code-loc-per-doc}{455.07}
\DefMacro{corpus-multinomials-num-docs}{5}
\DefMacro{corpus-multinomials-spec-loc}{3,699}
\DefMacro{corpus-multinomials-k-spec-loc}{3}
\DefMacro{corpus-multinomials-proof-loc}{3,664}
\DefMacro{corpus-multinomials-k-proof-loc}{3}
\DefMacro{corpus-multinomials-comment-loc}{308}
\DefMacro{corpus-multinomials-k-comment-loc}{0}
\DefMacro{corpus-multinomials-total-loc}{7,671}
\DefMacro{corpus-multinomials-k-total-loc}{7}
\DefMacro{corpus-multinomials-code-loc}{7,363}
\DefMacro{corpus-multinomials-k-code-loc}{7}
\DefMacro{corpus-multinomials-AVG-spec-loc-per-doc}{739.80}
\DefMacro{corpus-multinomials-SUM-spec-loc-per-doc}{3,699}
\DefMacro{corpus-multinomials-MAX-spec-loc-per-doc}{2,069}
\DefMacro{corpus-multinomials-MIN-spec-loc-per-doc}{23}
\DefMacro{corpus-multinomials-MEDIAN-spec-loc-per-doc}{502.00}
\DefMacro{corpus-multinomials-STDEV-spec-loc-per-doc}{737.37}
\DefMacro{corpus-multinomials-AVG-proof-loc-per-doc}{732.80}
\DefMacro{corpus-multinomials-SUM-proof-loc-per-doc}{3,664}
\DefMacro{corpus-multinomials-MAX-proof-loc-per-doc}{2,387}
\DefMacro{corpus-multinomials-MIN-proof-loc-per-doc}{5}
\DefMacro{corpus-multinomials-MEDIAN-proof-loc-per-doc}{521.00}
\DefMacro{corpus-multinomials-STDEV-proof-loc-per-doc}{855.37}
\DefMacro{corpus-multinomials-AVG-comment-loc-per-doc}{61.60}
\DefMacro{corpus-multinomials-SUM-comment-loc-per-doc}{308}
\DefMacro{corpus-multinomials-MAX-comment-loc-per-doc}{155}
\DefMacro{corpus-multinomials-MIN-comment-loc-per-doc}{2}
\DefMacro{corpus-multinomials-MEDIAN-comment-loc-per-doc}{51.00}
\DefMacro{corpus-multinomials-STDEV-comment-loc-per-doc}{53.55}
\DefMacro{corpus-multinomials-AVG-total-loc-per-doc}{1,534.20}
\DefMacro{corpus-multinomials-SUM-total-loc-per-doc}{7,671}
\DefMacro{corpus-multinomials-MAX-total-loc-per-doc}{4,611}
\DefMacro{corpus-multinomials-MIN-total-loc-per-doc}{30}
\DefMacro{corpus-multinomials-MEDIAN-total-loc-per-doc}{1,143.00}
\DefMacro{corpus-multinomials-STDEV-total-loc-per-doc}{1,631.48}
\DefMacro{corpus-multinomials-AVG-code-loc-per-doc}{1,472.60}
\DefMacro{corpus-multinomials-SUM-code-loc-per-doc}{7,363}
\DefMacro{corpus-multinomials-MAX-code-loc-per-doc}{4,456}
\DefMacro{corpus-multinomials-MIN-code-loc-per-doc}{28}
\DefMacro{corpus-multinomials-MEDIAN-code-loc-per-doc}{1,092.00}
\DefMacro{corpus-multinomials-STDEV-code-loc-per-doc}{1,578.81}
\DefMacro{corpus-ellipticcurves-num-docs}{18}
\DefMacro{corpus-ellipticcurves-spec-loc}{3,298}
\DefMacro{corpus-ellipticcurves-k-spec-loc}{3}
\DefMacro{corpus-ellipticcurves-proof-loc}{6,298}
\DefMacro{corpus-ellipticcurves-k-proof-loc}{6}
\DefMacro{corpus-ellipticcurves-comment-loc}{279}
\DefMacro{corpus-ellipticcurves-k-comment-loc}{0}
\DefMacro{corpus-ellipticcurves-total-loc}{9,875}
\DefMacro{corpus-ellipticcurves-k-total-loc}{9}
\DefMacro{corpus-ellipticcurves-code-loc}{9,596}
\DefMacro{corpus-ellipticcurves-k-code-loc}{9}
\DefMacro{corpus-ellipticcurves-AVG-spec-loc-per-doc}{183.22}
\DefMacro{corpus-ellipticcurves-SUM-spec-loc-per-doc}{3,298}
\DefMacro{corpus-ellipticcurves-MAX-spec-loc-per-doc}{503}
\DefMacro{corpus-ellipticcurves-MIN-spec-loc-per-doc}{38}
\DefMacro{corpus-ellipticcurves-MEDIAN-spec-loc-per-doc}{150.50}
\DefMacro{corpus-ellipticcurves-STDEV-spec-loc-per-doc}{136.29}
\DefMacro{corpus-ellipticcurves-AVG-proof-loc-per-doc}{349.89}
\DefMacro{corpus-ellipticcurves-SUM-proof-loc-per-doc}{6,298}
\DefMacro{corpus-ellipticcurves-MAX-proof-loc-per-doc}{1,115}
\DefMacro{corpus-ellipticcurves-MIN-proof-loc-per-doc}{48}
\DefMacro{corpus-ellipticcurves-MEDIAN-proof-loc-per-doc}{271.00}
\DefMacro{corpus-ellipticcurves-STDEV-proof-loc-per-doc}{278.67}
\DefMacro{corpus-ellipticcurves-AVG-comment-loc-per-doc}{15.50}
\DefMacro{corpus-ellipticcurves-SUM-comment-loc-per-doc}{279}
\DefMacro{corpus-ellipticcurves-MAX-comment-loc-per-doc}{36}
\DefMacro{corpus-ellipticcurves-MIN-comment-loc-per-doc}{2}
\DefMacro{corpus-ellipticcurves-MEDIAN-comment-loc-per-doc}{14.50}
\DefMacro{corpus-ellipticcurves-STDEV-comment-loc-per-doc}{9.12}
\DefMacro{corpus-ellipticcurves-AVG-total-loc-per-doc}{548.61}
\DefMacro{corpus-ellipticcurves-SUM-total-loc-per-doc}{9,875}
\DefMacro{corpus-ellipticcurves-MAX-total-loc-per-doc}{1,649}
\DefMacro{corpus-ellipticcurves-MIN-total-loc-per-doc}{100}
\DefMacro{corpus-ellipticcurves-MEDIAN-total-loc-per-doc}{400.50}
\DefMacro{corpus-ellipticcurves-STDEV-total-loc-per-doc}{388.61}
\DefMacro{corpus-ellipticcurves-AVG-code-loc-per-doc}{533.11}
\DefMacro{corpus-ellipticcurves-SUM-code-loc-per-doc}{9,596}
\DefMacro{corpus-ellipticcurves-MAX-code-loc-per-doc}{1,615}
\DefMacro{corpus-ellipticcurves-MIN-code-loc-per-doc}{94}
\DefMacro{corpus-ellipticcurves-MEDIAN-code-loc-per-doc}{390.00}
\DefMacro{corpus-ellipticcurves-STDEV-code-loc-per-doc}{384.65}
\DefMacro{corpus-toychain-num-docs}{14}
\DefMacro{corpus-toychain-spec-loc}{1,747}
\DefMacro{corpus-toychain-k-spec-loc}{1}
\DefMacro{corpus-toychain-proof-loc}{3,528}
\DefMacro{corpus-toychain-k-proof-loc}{3}
\DefMacro{corpus-toychain-comment-loc}{263}
\DefMacro{corpus-toychain-k-comment-loc}{0}
\DefMacro{corpus-toychain-total-loc}{5,538}
\DefMacro{corpus-toychain-k-total-loc}{5}
\DefMacro{corpus-toychain-code-loc}{5,275}
\DefMacro{corpus-toychain-k-code-loc}{5}
\DefMacro{corpus-toychain-AVG-spec-loc-per-doc}{124.79}
\DefMacro{corpus-toychain-SUM-spec-loc-per-doc}{1,747}
\DefMacro{corpus-toychain-MAX-spec-loc-per-doc}{567}
\DefMacro{corpus-toychain-MIN-spec-loc-per-doc}{10}
\DefMacro{corpus-toychain-MEDIAN-spec-loc-per-doc}{57.00}
\DefMacro{corpus-toychain-STDEV-spec-loc-per-doc}{149.06}
\DefMacro{corpus-toychain-AVG-proof-loc-per-doc}{252.00}
\DefMacro{corpus-toychain-SUM-proof-loc-per-doc}{3,528}
\DefMacro{corpus-toychain-MAX-proof-loc-per-doc}{2,176}
\DefMacro{corpus-toychain-MIN-proof-loc-per-doc}{0}
\DefMacro{corpus-toychain-MEDIAN-proof-loc-per-doc}{107.50}
\DefMacro{corpus-toychain-STDEV-proof-loc-per-doc}{546.07}
\DefMacro{corpus-toychain-AVG-comment-loc-per-doc}{18.79}
\DefMacro{corpus-toychain-SUM-comment-loc-per-doc}{263}
\DefMacro{corpus-toychain-MAX-comment-loc-per-doc}{78}
\DefMacro{corpus-toychain-MIN-comment-loc-per-doc}{1}
\DefMacro{corpus-toychain-MEDIAN-comment-loc-per-doc}{6.00}
\DefMacro{corpus-toychain-STDEV-comment-loc-per-doc}{25.46}
\DefMacro{corpus-toychain-AVG-total-loc-per-doc}{395.57}
\DefMacro{corpus-toychain-SUM-total-loc-per-doc}{5,538}
\DefMacro{corpus-toychain-MAX-total-loc-per-doc}{2,821}
\DefMacro{corpus-toychain-MIN-total-loc-per-doc}{21}
\DefMacro{corpus-toychain-MEDIAN-total-loc-per-doc}{167.00}
\DefMacro{corpus-toychain-STDEV-total-loc-per-doc}{705.51}
\DefMacro{corpus-toychain-AVG-code-loc-per-doc}{376.79}
\DefMacro{corpus-toychain-SUM-code-loc-per-doc}{5,275}
\DefMacro{corpus-toychain-MAX-code-loc-per-doc}{2,743}
\DefMacro{corpus-toychain-MIN-code-loc-per-doc}{14}
\DefMacro{corpus-toychain-MEDIAN-code-loc-per-doc}{164.00}
\DefMacro{corpus-toychain-STDEV-code-loc-per-doc}{685.03}
\DefMacro{corpus-bits-num-docs}{10}
\DefMacro{corpus-bits-spec-loc}{1,578}
\DefMacro{corpus-bits-k-spec-loc}{1}
\DefMacro{corpus-bits-proof-loc}{2,463}
\DefMacro{corpus-bits-k-proof-loc}{2}
\DefMacro{corpus-bits-comment-loc}{554}
\DefMacro{corpus-bits-k-comment-loc}{0}
\DefMacro{corpus-bits-total-loc}{4,595}
\DefMacro{corpus-bits-k-total-loc}{4}
\DefMacro{corpus-bits-code-loc}{4,041}
\DefMacro{corpus-bits-k-code-loc}{4}
\DefMacro{corpus-bits-AVG-spec-loc-per-doc}{157.80}
\DefMacro{corpus-bits-SUM-spec-loc-per-doc}{1,578}
\DefMacro{corpus-bits-MAX-spec-loc-per-doc}{407}
\DefMacro{corpus-bits-MIN-spec-loc-per-doc}{3}
\DefMacro{corpus-bits-MEDIAN-spec-loc-per-doc}{178.00}
\DefMacro{corpus-bits-STDEV-spec-loc-per-doc}{118.24}
\DefMacro{corpus-bits-AVG-proof-loc-per-doc}{246.30}
\DefMacro{corpus-bits-SUM-proof-loc-per-doc}{2,463}
\DefMacro{corpus-bits-MAX-proof-loc-per-doc}{1,204}
\DefMacro{corpus-bits-MIN-proof-loc-per-doc}{0}
\DefMacro{corpus-bits-MEDIAN-proof-loc-per-doc}{167.50}
\DefMacro{corpus-bits-STDEV-proof-loc-per-doc}{346.70}
\DefMacro{corpus-bits-AVG-comment-loc-per-doc}{55.40}
\DefMacro{corpus-bits-SUM-comment-loc-per-doc}{554}
\DefMacro{corpus-bits-MAX-comment-loc-per-doc}{188}
\DefMacro{corpus-bits-MIN-comment-loc-per-doc}{0}
\DefMacro{corpus-bits-MEDIAN-comment-loc-per-doc}{58.50}
\DefMacro{corpus-bits-STDEV-comment-loc-per-doc}{52.37}
\DefMacro{corpus-bits-AVG-total-loc-per-doc}{459.50}
\DefMacro{corpus-bits-SUM-total-loc-per-doc}{4,595}
\DefMacro{corpus-bits-MAX-total-loc-per-doc}{1,799}
\DefMacro{corpus-bits-MIN-total-loc-per-doc}{3}
\DefMacro{corpus-bits-MEDIAN-total-loc-per-doc}{411.00}
\DefMacro{corpus-bits-STDEV-total-loc-per-doc}{497.97}
\DefMacro{corpus-bits-AVG-code-loc-per-doc}{404.10}
\DefMacro{corpus-bits-SUM-code-loc-per-doc}{4,041}
\DefMacro{corpus-bits-MAX-code-loc-per-doc}{1,611}
\DefMacro{corpus-bits-MIN-code-loc-per-doc}{3}
\DefMacro{corpus-bits-MEDIAN-code-loc-per-doc}{354.50}
\DefMacro{corpus-bits-STDEV-code-loc-per-doc}{448.48}
\DefMacro{corpus-twosquare-num-docs}{2}
\DefMacro{corpus-twosquare-spec-loc}{413}
\DefMacro{corpus-twosquare-k-spec-loc}{0}
\DefMacro{corpus-twosquare-proof-loc}{1,308}
\DefMacro{corpus-twosquare-k-proof-loc}{1}
\DefMacro{corpus-twosquare-comment-loc}{78}
\DefMacro{corpus-twosquare-k-comment-loc}{0}
\DefMacro{corpus-twosquare-total-loc}{1,799}
\DefMacro{corpus-twosquare-k-total-loc}{1}
\DefMacro{corpus-twosquare-code-loc}{1,721}
\DefMacro{corpus-twosquare-k-code-loc}{1}
\DefMacro{corpus-twosquare-AVG-spec-loc-per-doc}{206.50}
\DefMacro{corpus-twosquare-SUM-spec-loc-per-doc}{413}
\DefMacro{corpus-twosquare-MAX-spec-loc-per-doc}{369}
\DefMacro{corpus-twosquare-MIN-spec-loc-per-doc}{44}
\DefMacro{corpus-twosquare-MEDIAN-spec-loc-per-doc}{206.50}
\DefMacro{corpus-twosquare-STDEV-spec-loc-per-doc}{162.50}
\DefMacro{corpus-twosquare-AVG-proof-loc-per-doc}{654.00}
\DefMacro{corpus-twosquare-SUM-proof-loc-per-doc}{1,308}
\DefMacro{corpus-twosquare-MAX-proof-loc-per-doc}{1,126}
\DefMacro{corpus-twosquare-MIN-proof-loc-per-doc}{182}
\DefMacro{corpus-twosquare-MEDIAN-proof-loc-per-doc}{654.00}
\DefMacro{corpus-twosquare-STDEV-proof-loc-per-doc}{472.00}
\DefMacro{corpus-twosquare-AVG-comment-loc-per-doc}{39.00}
\DefMacro{corpus-twosquare-SUM-comment-loc-per-doc}{78}
\DefMacro{corpus-twosquare-MAX-comment-loc-per-doc}{77}
\DefMacro{corpus-twosquare-MIN-comment-loc-per-doc}{1}
\DefMacro{corpus-twosquare-MEDIAN-comment-loc-per-doc}{39.00}
\DefMacro{corpus-twosquare-STDEV-comment-loc-per-doc}{38.00}
\DefMacro{corpus-twosquare-AVG-total-loc-per-doc}{899.50}
\DefMacro{corpus-twosquare-SUM-total-loc-per-doc}{1,799}
\DefMacro{corpus-twosquare-MAX-total-loc-per-doc}{1,572}
\DefMacro{corpus-twosquare-MIN-total-loc-per-doc}{227}
\DefMacro{corpus-twosquare-MEDIAN-total-loc-per-doc}{899.50}
\DefMacro{corpus-twosquare-STDEV-total-loc-per-doc}{672.50}
\DefMacro{corpus-twosquare-AVG-code-loc-per-doc}{860.50}
\DefMacro{corpus-twosquare-SUM-code-loc-per-doc}{1,721}
\DefMacro{corpus-twosquare-MAX-code-loc-per-doc}{1,495}
\DefMacro{corpus-twosquare-MIN-code-loc-per-doc}{226}
\DefMacro{corpus-twosquare-MEDIAN-code-loc-per-doc}{860.50}
\DefMacro{corpus-twosquare-STDEV-code-loc-per-doc}{634.50}
\DefMacro{corpus-grobner-num-docs}{1}
\DefMacro{corpus-grobner-spec-loc}{312}
\DefMacro{corpus-grobner-k-spec-loc}{0}
\DefMacro{corpus-grobner-proof-loc}{1,018}
\DefMacro{corpus-grobner-k-proof-loc}{1}
\DefMacro{corpus-grobner-comment-loc}{48}
\DefMacro{corpus-grobner-k-comment-loc}{0}
\DefMacro{corpus-grobner-total-loc}{1,378}
\DefMacro{corpus-grobner-k-total-loc}{1}
\DefMacro{corpus-grobner-code-loc}{1,330}
\DefMacro{corpus-grobner-k-code-loc}{1}
\DefMacro{corpus-grobner-AVG-spec-loc-per-doc}{312.00}
\DefMacro{corpus-grobner-SUM-spec-loc-per-doc}{312}
\DefMacro{corpus-grobner-MAX-spec-loc-per-doc}{312}
\DefMacro{corpus-grobner-MIN-spec-loc-per-doc}{312}
\DefMacro{corpus-grobner-MEDIAN-spec-loc-per-doc}{312.00}
\DefMacro{corpus-grobner-STDEV-spec-loc-per-doc}{0.00}
\DefMacro{corpus-grobner-AVG-proof-loc-per-doc}{1,018.00}
\DefMacro{corpus-grobner-SUM-proof-loc-per-doc}{1,018}
\DefMacro{corpus-grobner-MAX-proof-loc-per-doc}{1,018}
\DefMacro{corpus-grobner-MIN-proof-loc-per-doc}{1,018}
\DefMacro{corpus-grobner-MEDIAN-proof-loc-per-doc}{1,018.00}
\DefMacro{corpus-grobner-STDEV-proof-loc-per-doc}{0.00}
\DefMacro{corpus-grobner-AVG-comment-loc-per-doc}{48.00}
\DefMacro{corpus-grobner-SUM-comment-loc-per-doc}{48}
\DefMacro{corpus-grobner-MAX-comment-loc-per-doc}{48}
\DefMacro{corpus-grobner-MIN-comment-loc-per-doc}{48}
\DefMacro{corpus-grobner-MEDIAN-comment-loc-per-doc}{48.00}
\DefMacro{corpus-grobner-STDEV-comment-loc-per-doc}{0.00}
\DefMacro{corpus-grobner-AVG-total-loc-per-doc}{1,378.00}
\DefMacro{corpus-grobner-SUM-total-loc-per-doc}{1,378}
\DefMacro{corpus-grobner-MAX-total-loc-per-doc}{1,378}
\DefMacro{corpus-grobner-MIN-total-loc-per-doc}{1,378}
\DefMacro{corpus-grobner-MEDIAN-total-loc-per-doc}{1,378.00}
\DefMacro{corpus-grobner-STDEV-total-loc-per-doc}{0.00}
\DefMacro{corpus-grobner-AVG-code-loc-per-doc}{1,330.00}
\DefMacro{corpus-grobner-SUM-code-loc-per-doc}{1,330}
\DefMacro{corpus-grobner-MAX-code-loc-per-doc}{1,330}
\DefMacro{corpus-grobner-MIN-code-loc-per-doc}{1,330}
\DefMacro{corpus-grobner-MEDIAN-code-loc-per-doc}{1,330.00}
\DefMacro{corpus-grobner-STDEV-code-loc-per-doc}{0.00}
\DefMacro{corpus-infotheo-num-docs}{81}
\DefMacro{corpus-infotheo-spec-loc}{12,517}
\DefMacro{corpus-infotheo-k-spec-loc}{12}
\DefMacro{corpus-infotheo-proof-loc}{29,778}
\DefMacro{corpus-infotheo-k-proof-loc}{29}
\DefMacro{corpus-infotheo-comment-loc}{1,733}
\DefMacro{corpus-infotheo-k-comment-loc}{1}
\DefMacro{corpus-infotheo-total-loc}{44,028}
\DefMacro{corpus-infotheo-k-total-loc}{44}
\DefMacro{corpus-infotheo-code-loc}{42,295}
\DefMacro{corpus-infotheo-k-code-loc}{42}
\DefMacro{corpus-infotheo-AVG-spec-loc-per-doc}{154.53}
\DefMacro{corpus-infotheo-SUM-spec-loc-per-doc}{12,517}
\DefMacro{corpus-infotheo-MAX-spec-loc-per-doc}{1,265}
\DefMacro{corpus-infotheo-MIN-spec-loc-per-doc}{15}
\DefMacro{corpus-infotheo-MEDIAN-spec-loc-per-doc}{104.00}
\DefMacro{corpus-infotheo-STDEV-spec-loc-per-doc}{181.69}
\DefMacro{corpus-infotheo-AVG-proof-loc-per-doc}{367.63}
\DefMacro{corpus-infotheo-SUM-proof-loc-per-doc}{29,778}
\DefMacro{corpus-infotheo-MAX-proof-loc-per-doc}{2,963}
\DefMacro{corpus-infotheo-MIN-proof-loc-per-doc}{0}
\DefMacro{corpus-infotheo-MEDIAN-proof-loc-per-doc}{235.00}
\DefMacro{corpus-infotheo-STDEV-proof-loc-per-doc}{430.38}
\DefMacro{corpus-infotheo-AVG-comment-loc-per-doc}{21.40}
\DefMacro{corpus-infotheo-SUM-comment-loc-per-doc}{1,733}
\DefMacro{corpus-infotheo-MAX-comment-loc-per-doc}{320}
\DefMacro{corpus-infotheo-MIN-comment-loc-per-doc}{0}
\DefMacro{corpus-infotheo-MEDIAN-comment-loc-per-doc}{13.00}
\DefMacro{corpus-infotheo-STDEV-comment-loc-per-doc}{39.29}
\DefMacro{corpus-infotheo-AVG-total-loc-per-doc}{543.56}
\DefMacro{corpus-infotheo-SUM-total-loc-per-doc}{44,028}
\DefMacro{corpus-infotheo-MAX-total-loc-per-doc}{4,548}
\DefMacro{corpus-infotheo-MIN-total-loc-per-doc}{33}
\DefMacro{corpus-infotheo-MEDIAN-total-loc-per-doc}{342.00}
\DefMacro{corpus-infotheo-STDEV-total-loc-per-doc}{625.20}
\DefMacro{corpus-infotheo-AVG-code-loc-per-doc}{522.16}
\DefMacro{corpus-infotheo-SUM-code-loc-per-doc}{42,295}
\DefMacro{corpus-infotheo-MAX-code-loc-per-doc}{4,228}
\DefMacro{corpus-infotheo-MIN-code-loc-per-doc}{31}
\DefMacro{corpus-infotheo-MEDIAN-code-loc-per-doc}{327.00}
\DefMacro{corpus-infotheo-STDEV-code-loc-per-doc}{591.23}
\DefMacro{corpus-monae-num-docs}{18}
\DefMacro{corpus-monae-spec-loc}{3,422}
\DefMacro{corpus-monae-k-spec-loc}{3}
\DefMacro{corpus-monae-proof-loc}{3,233}
\DefMacro{corpus-monae-k-proof-loc}{3}
\DefMacro{corpus-monae-comment-loc}{414}
\DefMacro{corpus-monae-k-comment-loc}{0}
\DefMacro{corpus-monae-total-loc}{7,069}
\DefMacro{corpus-monae-k-total-loc}{7}
\DefMacro{corpus-monae-code-loc}{6,655}
\DefMacro{corpus-monae-k-code-loc}{6}
\DefMacro{corpus-monae-AVG-spec-loc-per-doc}{190.11}
\DefMacro{corpus-monae-SUM-spec-loc-per-doc}{3,422}
\DefMacro{corpus-monae-MAX-spec-loc-per-doc}{844}
\DefMacro{corpus-monae-MIN-spec-loc-per-doc}{9}
\DefMacro{corpus-monae-MEDIAN-spec-loc-per-doc}{117.00}
\DefMacro{corpus-monae-STDEV-spec-loc-per-doc}{200.22}
\DefMacro{corpus-monae-AVG-proof-loc-per-doc}{179.61}
\DefMacro{corpus-monae-SUM-proof-loc-per-doc}{3,233}
\DefMacro{corpus-monae-MAX-proof-loc-per-doc}{451}
\DefMacro{corpus-monae-MIN-proof-loc-per-doc}{0}
\DefMacro{corpus-monae-MEDIAN-proof-loc-per-doc}{185.00}
\DefMacro{corpus-monae-STDEV-proof-loc-per-doc}{144.72}
\DefMacro{corpus-monae-AVG-comment-loc-per-doc}{23.00}
\DefMacro{corpus-monae-SUM-comment-loc-per-doc}{414}
\DefMacro{corpus-monae-MAX-comment-loc-per-doc}{161}
\DefMacro{corpus-monae-MIN-comment-loc-per-doc}{0}
\DefMacro{corpus-monae-MEDIAN-comment-loc-per-doc}{10.00}
\DefMacro{corpus-monae-STDEV-comment-loc-per-doc}{37.49}
\DefMacro{corpus-monae-AVG-total-loc-per-doc}{392.72}
\DefMacro{corpus-monae-SUM-total-loc-per-doc}{7,069}
\DefMacro{corpus-monae-MAX-total-loc-per-doc}{1,456}
\DefMacro{corpus-monae-MIN-total-loc-per-doc}{9}
\DefMacro{corpus-monae-MEDIAN-total-loc-per-doc}{368.50}
\DefMacro{corpus-monae-STDEV-total-loc-per-doc}{344.59}
\DefMacro{corpus-monae-AVG-code-loc-per-doc}{369.72}
\DefMacro{corpus-monae-SUM-code-loc-per-doc}{6,655}
\DefMacro{corpus-monae-MAX-code-loc-per-doc}{1,295}
\DefMacro{corpus-monae-MIN-code-loc-per-doc}{9}
\DefMacro{corpus-monae-MEDIAN-code-loc-per-doc}{366.00}
\DefMacro{corpus-monae-STDEV-code-loc-per-doc}{311.19}
\DefMacro{corpus-games-num-docs}{12}
\DefMacro{corpus-games-spec-loc}{1,450}
\DefMacro{corpus-games-k-spec-loc}{1}
\DefMacro{corpus-games-proof-loc}{3,503}
\DefMacro{corpus-games-k-proof-loc}{3}
\DefMacro{corpus-games-comment-loc}{258}
\DefMacro{corpus-games-k-comment-loc}{0}
\DefMacro{corpus-games-total-loc}{5,211}
\DefMacro{corpus-games-k-total-loc}{5}
\DefMacro{corpus-games-code-loc}{4,953}
\DefMacro{corpus-games-k-code-loc}{4}
\DefMacro{corpus-games-AVG-spec-loc-per-doc}{120.83}
\DefMacro{corpus-games-SUM-spec-loc-per-doc}{1,450}
\DefMacro{corpus-games-MAX-spec-loc-per-doc}{300}
\DefMacro{corpus-games-MIN-spec-loc-per-doc}{28}
\DefMacro{corpus-games-MEDIAN-spec-loc-per-doc}{99.00}
\DefMacro{corpus-games-STDEV-spec-loc-per-doc}{78.32}
\DefMacro{corpus-games-AVG-proof-loc-per-doc}{291.92}
\DefMacro{corpus-games-SUM-proof-loc-per-doc}{3,503}
\DefMacro{corpus-games-MAX-proof-loc-per-doc}{1,163}
\DefMacro{corpus-games-MIN-proof-loc-per-doc}{37}
\DefMacro{corpus-games-MEDIAN-proof-loc-per-doc}{184.00}
\DefMacro{corpus-games-STDEV-proof-loc-per-doc}{301.35}
\DefMacro{corpus-games-AVG-comment-loc-per-doc}{21.50}
\DefMacro{corpus-games-SUM-comment-loc-per-doc}{258}
\DefMacro{corpus-games-MAX-comment-loc-per-doc}{108}
\DefMacro{corpus-games-MIN-comment-loc-per-doc}{1}
\DefMacro{corpus-games-MEDIAN-comment-loc-per-doc}{14.00}
\DefMacro{corpus-games-STDEV-comment-loc-per-doc}{27.74}
\DefMacro{corpus-games-AVG-total-loc-per-doc}{434.25}
\DefMacro{corpus-games-SUM-total-loc-per-doc}{5,211}
\DefMacro{corpus-games-MAX-total-loc-per-doc}{1,483}
\DefMacro{corpus-games-MIN-total-loc-per-doc}{66}
\DefMacro{corpus-games-MEDIAN-total-loc-per-doc}{303.50}
\DefMacro{corpus-games-STDEV-total-loc-per-doc}{380.89}
\DefMacro{corpus-games-AVG-code-loc-per-doc}{412.75}
\DefMacro{corpus-games-SUM-code-loc-per-doc}{4,953}
\DefMacro{corpus-games-MAX-code-loc-per-doc}{1,463}
\DefMacro{corpus-games-MIN-code-loc-per-doc}{65}
\DefMacro{corpus-games-MEDIAN-code-loc-per-doc}{281.00}
\DefMacro{corpus-games-STDEV-code-loc-per-doc}{370.26}
\DefMacro{corpus-t1-num-projects}{4}
\DefMacro{corpus-t1-AVG-num-docs}{46.75}
\DefMacro{corpus-t1-AVG-spec-loc}{15,890.00}
\DefMacro{corpus-t1-AVG-proof-loc}{25,216.75}
\DefMacro{corpus-t1-AVG-comment-loc}{2,874.50}
\DefMacro{corpus-t1-AVG-total-loc}{43,981.25}
\DefMacro{corpus-t1-AVG-code-loc}{41,106.75}
\DefMacro{corpus-t1-SUM-num-docs}{187}
\DefMacro{corpus-t1-SUM-spec-loc}{63,560}
\DefMacro{corpus-t1-SUM-k-spec-loc}{63}
\DefMacro{corpus-t1-SUM-proof-loc}{100,867}
\DefMacro{corpus-t1-SUM-k-proof-loc}{100}
\DefMacro{corpus-t1-SUM-comment-loc}{11,498}
\DefMacro{corpus-t1-SUM-k-comment-loc}{11}
\DefMacro{corpus-t1-SUM-total-loc}{175,925}
\DefMacro{corpus-t1-SUM-k-total-loc}{175}
\DefMacro{corpus-t1-SUM-code-loc}{164,427}
\DefMacro{corpus-t1-SUM-k-code-loc}{164}
\DefMacro{corpus-t1-MAX-num-docs}{89}
\DefMacro{corpus-t1-MAX-spec-loc}{38,243}
\DefMacro{corpus-t1-MAX-k-spec-loc}{38}
\DefMacro{corpus-t1-MAX-proof-loc}{46,470}
\DefMacro{corpus-t1-MAX-k-proof-loc}{46}
\DefMacro{corpus-t1-MAX-comment-loc}{6,131}
\DefMacro{corpus-t1-MAX-k-comment-loc}{6}
\DefMacro{corpus-t1-MAX-total-loc}{90,844}
\DefMacro{corpus-t1-MAX-k-total-loc}{90}
\DefMacro{corpus-t1-MAX-code-loc}{84,713}
\DefMacro{corpus-t1-MAX-k-code-loc}{84}
\DefMacro{corpus-t1-MIN-num-docs}{4}
\DefMacro{corpus-t1-MIN-spec-loc}{4,260}
\DefMacro{corpus-t1-MIN-k-spec-loc}{4}
\DefMacro{corpus-t1-MIN-proof-loc}{2,191}
\DefMacro{corpus-t1-MIN-k-proof-loc}{2}
\DefMacro{corpus-t1-MIN-comment-loc}{238}
\DefMacro{corpus-t1-MIN-k-comment-loc}{0}
\DefMacro{corpus-t1-MIN-total-loc}{6,689}
\DefMacro{corpus-t1-MIN-k-total-loc}{6}
\DefMacro{corpus-t1-MIN-code-loc}{6,451}
\DefMacro{corpus-t1-MIN-k-code-loc}{6}
\DefMacro{corpus-t1-MEDIAN-num-docs}{47.00}
\DefMacro{corpus-t1-MEDIAN-spec-loc}{10,528.50}
\DefMacro{corpus-t1-MEDIAN-proof-loc}{26,103.00}
\DefMacro{corpus-t1-MEDIAN-comment-loc}{2,564.50}
\DefMacro{corpus-t1-MEDIAN-total-loc}{39,196.00}
\DefMacro{corpus-t1-MEDIAN-code-loc}{36,631.50}
\DefMacro{corpus-t1-STDEV-num-docs}{31.43}
\DefMacro{corpus-t1-STDEV-spec-loc}{13,191.56}
\DefMacro{corpus-t1-STDEV-proof-loc}{15,735.12}
\DefMacro{corpus-t1-STDEV-comment-loc}{2,113.92}
\DefMacro{corpus-t1-STDEV-total-loc}{30,140.40}
\DefMacro{corpus-t1-STDEV-code-loc}{28,031.66}
\DefMacro{corpus-t1-AVG-spec-loc-per-doc}{339.89}
\DefMacro{corpus-t1-SUM-spec-loc-per-doc}{63,560}
\DefMacro{corpus-t1-MAX-spec-loc-per-doc}{4,331}
\DefMacro{corpus-t1-MIN-spec-loc-per-doc}{1}
\DefMacro{corpus-t1-MEDIAN-spec-loc-per-doc}{185.00}
\DefMacro{corpus-t1-STDEV-spec-loc-per-doc}{462.97}
\DefMacro{corpus-t1-AVG-proof-loc-per-doc}{539.40}
\DefMacro{corpus-t1-SUM-proof-loc-per-doc}{100,867}
\DefMacro{corpus-t1-MAX-proof-loc-per-doc}{5,551}
\DefMacro{corpus-t1-MIN-proof-loc-per-doc}{0}
\DefMacro{corpus-t1-MEDIAN-proof-loc-per-doc}{381.00}
\DefMacro{corpus-t1-STDEV-proof-loc-per-doc}{704.12}
\DefMacro{corpus-t1-AVG-comment-loc-per-doc}{61.49}
\DefMacro{corpus-t1-SUM-comment-loc-per-doc}{11,498}
\DefMacro{corpus-t1-MAX-comment-loc-per-doc}{588}
\DefMacro{corpus-t1-MIN-comment-loc-per-doc}{0}
\DefMacro{corpus-t1-MEDIAN-comment-loc-per-doc}{43.00}
\DefMacro{corpus-t1-STDEV-comment-loc-per-doc}{66.75}
\DefMacro{corpus-t1-AVG-total-loc-per-doc}{940.78}
\DefMacro{corpus-t1-SUM-total-loc-per-doc}{175,925}
\DefMacro{corpus-t1-MAX-total-loc-per-doc}{5,619}
\DefMacro{corpus-t1-MIN-total-loc-per-doc}{1}
\DefMacro{corpus-t1-MEDIAN-total-loc-per-doc}{654.00}
\DefMacro{corpus-t1-STDEV-total-loc-per-doc}{964.69}
\DefMacro{corpus-t1-AVG-code-loc-per-doc}{879.29}
\DefMacro{corpus-t1-SUM-code-loc-per-doc}{164,427}
\DefMacro{corpus-t1-MAX-code-loc-per-doc}{5,560}
\DefMacro{corpus-t1-MIN-code-loc-per-doc}{1}
\DefMacro{corpus-t1-MEDIAN-code-loc-per-doc}{622.00}
\DefMacro{corpus-t1-STDEV-code-loc-per-doc}{925.44}
\DefMacro{corpus-t2-num-projects}{8}
\DefMacro{corpus-t2-AVG-num-docs}{8.38}
\DefMacro{corpus-t2-AVG-spec-loc}{2,696.75}
\DefMacro{corpus-t2-AVG-proof-loc}{3,788.50}
\DefMacro{corpus-t2-AVG-comment-loc}{385.62}
\DefMacro{corpus-t2-AVG-total-loc}{6,870.88}
\DefMacro{corpus-t2-AVG-code-loc}{6,485.25}
\DefMacro{corpus-t2-SUM-num-docs}{67}
\DefMacro{corpus-t2-SUM-spec-loc}{21,574}
\DefMacro{corpus-t2-SUM-k-spec-loc}{21}
\DefMacro{corpus-t2-SUM-proof-loc}{30,308}
\DefMacro{corpus-t2-SUM-k-proof-loc}{30}
\DefMacro{corpus-t2-SUM-comment-loc}{3,085}
\DefMacro{corpus-t2-SUM-k-comment-loc}{3}
\DefMacro{corpus-t2-SUM-total-loc}{54,967}
\DefMacro{corpus-t2-SUM-k-total-loc}{54}
\DefMacro{corpus-t2-SUM-code-loc}{51,882}
\DefMacro{corpus-t2-SUM-k-code-loc}{51}
\DefMacro{corpus-t2-MAX-num-docs}{18}
\DefMacro{corpus-t2-MAX-spec-loc}{5,553}
\DefMacro{corpus-t2-MAX-k-spec-loc}{5}
\DefMacro{corpus-t2-MAX-proof-loc}{7,249}
\DefMacro{corpus-t2-MAX-k-proof-loc}{7}
\DefMacro{corpus-t2-MAX-comment-loc}{1,352}
\DefMacro{corpus-t2-MAX-k-comment-loc}{1}
\DefMacro{corpus-t2-MAX-total-loc}{13,091}
\DefMacro{corpus-t2-MAX-k-total-loc}{13}
\DefMacro{corpus-t2-MAX-code-loc}{11,739}
\DefMacro{corpus-t2-MAX-k-code-loc}{11}
\DefMacro{corpus-t2-MIN-num-docs}{1}
\DefMacro{corpus-t2-MIN-spec-loc}{70}
\DefMacro{corpus-t2-MIN-k-spec-loc}{0}
\DefMacro{corpus-t2-MIN-proof-loc}{8}
\DefMacro{corpus-t2-MIN-k-proof-loc}{0}
\DefMacro{corpus-t2-MIN-comment-loc}{16}
\DefMacro{corpus-t2-MIN-k-comment-loc}{0}
\DefMacro{corpus-t2-MIN-total-loc}{94}
\DefMacro{corpus-t2-MIN-k-total-loc}{0}
\DefMacro{corpus-t2-MIN-code-loc}{78}
\DefMacro{corpus-t2-MIN-k-code-loc}{0}
\DefMacro{corpus-t2-MEDIAN-num-docs}{7.50}
\DefMacro{corpus-t2-MEDIAN-spec-loc}{3,498.50}
\DefMacro{corpus-t2-MEDIAN-proof-loc}{4,120.50}
\DefMacro{corpus-t2-MEDIAN-comment-loc}{293.50}
\DefMacro{corpus-t2-MEDIAN-total-loc}{8,476.00}
\DefMacro{corpus-t2-MEDIAN-code-loc}{8,144.00}
\DefMacro{corpus-t2-STDEV-num-docs}{6.63}
\DefMacro{corpus-t2-STDEV-spec-loc}{1,982.91}
\DefMacro{corpus-t2-STDEV-proof-loc}{2,568.53}
\DefMacro{corpus-t2-STDEV-comment-loc}{413.26}
\DefMacro{corpus-t2-STDEV-total-loc}{4,744.27}
\DefMacro{corpus-t2-STDEV-code-loc}{4,414.50}
\DefMacro{corpus-t2-AVG-spec-loc-per-doc}{322.00}
\DefMacro{corpus-t2-SUM-spec-loc-per-doc}{21,574}
\DefMacro{corpus-t2-MAX-spec-loc-per-doc}{2,069}
\DefMacro{corpus-t2-MIN-spec-loc-per-doc}{8}
\DefMacro{corpus-t2-MEDIAN-spec-loc-per-doc}{226.00}
\DefMacro{corpus-t2-STDEV-spec-loc-per-doc}{334.27}
\DefMacro{corpus-t2-AVG-proof-loc-per-doc}{452.36}
\DefMacro{corpus-t2-SUM-proof-loc-per-doc}{30,308}
\DefMacro{corpus-t2-MAX-proof-loc-per-doc}{2,387}
\DefMacro{corpus-t2-MIN-proof-loc-per-doc}{0}
\DefMacro{corpus-t2-MEDIAN-proof-loc-per-doc}{327.00}
\DefMacro{corpus-t2-STDEV-proof-loc-per-doc}{422.40}
\DefMacro{corpus-t2-AVG-comment-loc-per-doc}{46.04}
\DefMacro{corpus-t2-SUM-comment-loc-per-doc}{3,085}
\DefMacro{corpus-t2-MAX-comment-loc-per-doc}{311}
\DefMacro{corpus-t2-MIN-comment-loc-per-doc}{0}
\DefMacro{corpus-t2-MEDIAN-comment-loc-per-doc}{27.00}
\DefMacro{corpus-t2-STDEV-comment-loc-per-doc}{58.59}
\DefMacro{corpus-t2-AVG-total-loc-per-doc}{820.40}
\DefMacro{corpus-t2-SUM-total-loc-per-doc}{54,967}
\DefMacro{corpus-t2-MAX-total-loc-per-doc}{4,611}
\DefMacro{corpus-t2-MIN-total-loc-per-doc}{8}
\DefMacro{corpus-t2-MEDIAN-total-loc-per-doc}{747.00}
\DefMacro{corpus-t2-STDEV-total-loc-per-doc}{717.19}
\DefMacro{corpus-t2-AVG-code-loc-per-doc}{774.36}
\DefMacro{corpus-t2-SUM-code-loc-per-doc}{51,882}
\DefMacro{corpus-t2-MAX-code-loc-per-doc}{4,456}
\DefMacro{corpus-t2-MIN-code-loc-per-doc}{8}
\DefMacro{corpus-t2-MEDIAN-code-loc-per-doc}{720.00}
\DefMacro{corpus-t2-STDEV-code-loc-per-doc}{682.33}
\DefMacro{corpus-t3-num-projects}{8}
\DefMacro{corpus-t3-AVG-num-docs}{14.25}
\DefMacro{corpus-t3-AVG-spec-loc}{2,164.12}
\DefMacro{corpus-t3-AVG-proof-loc}{2,665.62}
\DefMacro{corpus-t3-AVG-comment-loc}{388.00}
\DefMacro{corpus-t3-AVG-total-loc}{5,217.75}
\DefMacro{corpus-t3-AVG-code-loc}{4,829.75}
\DefMacro{corpus-t3-SUM-num-docs}{114}
\DefMacro{corpus-t3-SUM-spec-loc}{17,313}
\DefMacro{corpus-t3-SUM-k-spec-loc}{17}
\DefMacro{corpus-t3-SUM-proof-loc}{21,325}
\DefMacro{corpus-t3-SUM-k-proof-loc}{21}
\DefMacro{corpus-t3-SUM-comment-loc}{3,104}
\DefMacro{corpus-t3-SUM-k-comment-loc}{3}
\DefMacro{corpus-t3-SUM-total-loc}{41,742}
\DefMacro{corpus-t3-SUM-k-total-loc}{41}
\DefMacro{corpus-t3-SUM-code-loc}{38,638}
\DefMacro{corpus-t3-SUM-k-code-loc}{38}
\DefMacro{corpus-t3-MAX-num-docs}{20}
\DefMacro{corpus-t3-MAX-spec-loc}{3,422}
\DefMacro{corpus-t3-MAX-k-spec-loc}{3}
\DefMacro{corpus-t3-MAX-proof-loc}{3,528}
\DefMacro{corpus-t3-MAX-k-proof-loc}{3}
\DefMacro{corpus-t3-MAX-comment-loc}{554}
\DefMacro{corpus-t3-MAX-k-comment-loc}{0}
\DefMacro{corpus-t3-MAX-total-loc}{7,069}
\DefMacro{corpus-t3-MAX-k-total-loc}{7}
\DefMacro{corpus-t3-MAX-code-loc}{6,655}
\DefMacro{corpus-t3-MAX-k-code-loc}{6}
\DefMacro{corpus-t3-MIN-num-docs}{10}
\DefMacro{corpus-t3-MIN-spec-loc}{1,299}
\DefMacro{corpus-t3-MIN-k-spec-loc}{1}
\DefMacro{corpus-t3-MIN-proof-loc}{1,734}
\DefMacro{corpus-t3-MIN-k-proof-loc}{1}
\DefMacro{corpus-t3-MIN-comment-loc}{250}
\DefMacro{corpus-t3-MIN-k-comment-loc}{0}
\DefMacro{corpus-t3-MIN-total-loc}{3,283}
\DefMacro{corpus-t3-MIN-k-total-loc}{3}
\DefMacro{corpus-t3-MIN-code-loc}{3,033}
\DefMacro{corpus-t3-MIN-k-code-loc}{3}
\DefMacro{corpus-t3-MEDIAN-num-docs}{13.00}
\DefMacro{corpus-t3-MEDIAN-spec-loc}{2,026.00}
\DefMacro{corpus-t3-MEDIAN-proof-loc}{2,657.50}
\DefMacro{corpus-t3-MEDIAN-comment-loc}{404.00}
\DefMacro{corpus-t3-MEDIAN-total-loc}{5,094.50}
\DefMacro{corpus-t3-MEDIAN-code-loc}{4,713.00}
\DefMacro{corpus-t3-STDEV-num-docs}{3.23}
\DefMacro{corpus-t3-STDEV-spec-loc}{719.74}
\DefMacro{corpus-t3-STDEV-proof-loc}{671.20}
\DefMacro{corpus-t3-STDEV-comment-loc}{111.59}
\DefMacro{corpus-t3-STDEV-total-loc}{1,058.18}
\DefMacro{corpus-t3-STDEV-code-loc}{1,037.52}
\DefMacro{corpus-t3-AVG-spec-loc-per-doc}{151.87}
\DefMacro{corpus-t3-SUM-spec-loc-per-doc}{17,313}
\DefMacro{corpus-t3-MAX-spec-loc-per-doc}{844}
\DefMacro{corpus-t3-MIN-spec-loc-per-doc}{3}
\DefMacro{corpus-t3-MEDIAN-spec-loc-per-doc}{103.00}
\DefMacro{corpus-t3-STDEV-spec-loc-per-doc}{148.08}
\DefMacro{corpus-t3-AVG-proof-loc-per-doc}{187.06}
\DefMacro{corpus-t3-SUM-proof-loc-per-doc}{21,325}
\DefMacro{corpus-t3-MAX-proof-loc-per-doc}{2,176}
\DefMacro{corpus-t3-MIN-proof-loc-per-doc}{0}
\DefMacro{corpus-t3-MEDIAN-proof-loc-per-doc}{103.50}
\DefMacro{corpus-t3-STDEV-proof-loc-per-doc}{280.20}
\DefMacro{corpus-t3-AVG-comment-loc-per-doc}{27.23}
\DefMacro{corpus-t3-SUM-comment-loc-per-doc}{3,104}
\DefMacro{corpus-t3-MAX-comment-loc-per-doc}{188}
\DefMacro{corpus-t3-MIN-comment-loc-per-doc}{0}
\DefMacro{corpus-t3-MEDIAN-comment-loc-per-doc}{17.00}
\DefMacro{corpus-t3-STDEV-comment-loc-per-doc}{32.96}
\DefMacro{corpus-t3-AVG-total-loc-per-doc}{366.16}
\DefMacro{corpus-t3-SUM-total-loc-per-doc}{41,742}
\DefMacro{corpus-t3-MAX-total-loc-per-doc}{2,821}
\DefMacro{corpus-t3-MIN-total-loc-per-doc}{3}
\DefMacro{corpus-t3-MEDIAN-total-loc-per-doc}{243.50}
\DefMacro{corpus-t3-STDEV-total-loc-per-doc}{413.48}
\DefMacro{corpus-t3-AVG-code-loc-per-doc}{338.93}
\DefMacro{corpus-t3-SUM-code-loc-per-doc}{38,638}
\DefMacro{corpus-t3-MAX-code-loc-per-doc}{2,743}
\DefMacro{corpus-t3-MIN-code-loc-per-doc}{3}
\DefMacro{corpus-t3-MEDIAN-code-loc-per-doc}{216.00}
\DefMacro{corpus-t3-STDEV-code-loc-per-doc}{390.96}
\DefMacro{corpus-lo-num-projects}{1}
\DefMacro{corpus-lo-AVG-num-docs}{81.00}
\DefMacro{corpus-lo-AVG-spec-loc}{12,517.00}
\DefMacro{corpus-lo-AVG-proof-loc}{29,778.00}
\DefMacro{corpus-lo-AVG-comment-loc}{1,733.00}
\DefMacro{corpus-lo-AVG-total-loc}{44,028.00}
\DefMacro{corpus-lo-AVG-code-loc}{42,295.00}
\DefMacro{corpus-lo-SUM-num-docs}{81}
\DefMacro{corpus-lo-SUM-spec-loc}{12,517}
\DefMacro{corpus-lo-SUM-k-spec-loc}{12}
\DefMacro{corpus-lo-SUM-proof-loc}{29,778}
\DefMacro{corpus-lo-SUM-k-proof-loc}{29}
\DefMacro{corpus-lo-SUM-comment-loc}{1,733}
\DefMacro{corpus-lo-SUM-k-comment-loc}{1}
\DefMacro{corpus-lo-SUM-total-loc}{44,028}
\DefMacro{corpus-lo-SUM-k-total-loc}{44}
\DefMacro{corpus-lo-SUM-code-loc}{42,295}
\DefMacro{corpus-lo-SUM-k-code-loc}{42}
\DefMacro{corpus-lo-MAX-num-docs}{81}
\DefMacro{corpus-lo-MAX-spec-loc}{12,517}
\DefMacro{corpus-lo-MAX-k-spec-loc}{12}
\DefMacro{corpus-lo-MAX-proof-loc}{29,778}
\DefMacro{corpus-lo-MAX-k-proof-loc}{29}
\DefMacro{corpus-lo-MAX-comment-loc}{1,733}
\DefMacro{corpus-lo-MAX-k-comment-loc}{1}
\DefMacro{corpus-lo-MAX-total-loc}{44,028}
\DefMacro{corpus-lo-MAX-k-total-loc}{44}
\DefMacro{corpus-lo-MAX-code-loc}{42,295}
\DefMacro{corpus-lo-MAX-k-code-loc}{42}
\DefMacro{corpus-lo-MIN-num-docs}{81}
\DefMacro{corpus-lo-MIN-spec-loc}{12,517}
\DefMacro{corpus-lo-MIN-k-spec-loc}{12}
\DefMacro{corpus-lo-MIN-proof-loc}{29,778}
\DefMacro{corpus-lo-MIN-k-proof-loc}{29}
\DefMacro{corpus-lo-MIN-comment-loc}{1,733}
\DefMacro{corpus-lo-MIN-k-comment-loc}{1}
\DefMacro{corpus-lo-MIN-total-loc}{44,028}
\DefMacro{corpus-lo-MIN-k-total-loc}{44}
\DefMacro{corpus-lo-MIN-code-loc}{42,295}
\DefMacro{corpus-lo-MIN-k-code-loc}{42}
\DefMacro{corpus-lo-MEDIAN-num-docs}{81.00}
\DefMacro{corpus-lo-MEDIAN-spec-loc}{12,517.00}
\DefMacro{corpus-lo-MEDIAN-proof-loc}{29,778.00}
\DefMacro{corpus-lo-MEDIAN-comment-loc}{1,733.00}
\DefMacro{corpus-lo-MEDIAN-total-loc}{44,028.00}
\DefMacro{corpus-lo-MEDIAN-code-loc}{42,295.00}
\DefMacro{corpus-lo-STDEV-num-docs}{0.00}
\DefMacro{corpus-lo-STDEV-spec-loc}{0.00}
\DefMacro{corpus-lo-STDEV-proof-loc}{0.00}
\DefMacro{corpus-lo-STDEV-comment-loc}{0.00}
\DefMacro{corpus-lo-STDEV-total-loc}{0.00}
\DefMacro{corpus-lo-STDEV-code-loc}{0.00}
\DefMacro{corpus-lo-AVG-spec-loc-per-doc}{154.53}
\DefMacro{corpus-lo-SUM-spec-loc-per-doc}{12,517}
\DefMacro{corpus-lo-MAX-spec-loc-per-doc}{1,265}
\DefMacro{corpus-lo-MIN-spec-loc-per-doc}{15}
\DefMacro{corpus-lo-MEDIAN-spec-loc-per-doc}{104.00}
\DefMacro{corpus-lo-STDEV-spec-loc-per-doc}{181.69}
\DefMacro{corpus-lo-AVG-proof-loc-per-doc}{367.63}
\DefMacro{corpus-lo-SUM-proof-loc-per-doc}{29,778}
\DefMacro{corpus-lo-MAX-proof-loc-per-doc}{2,963}
\DefMacro{corpus-lo-MIN-proof-loc-per-doc}{0}
\DefMacro{corpus-lo-MEDIAN-proof-loc-per-doc}{235.00}
\DefMacro{corpus-lo-STDEV-proof-loc-per-doc}{430.38}
\DefMacro{corpus-lo-AVG-comment-loc-per-doc}{21.40}
\DefMacro{corpus-lo-SUM-comment-loc-per-doc}{1,733}
\DefMacro{corpus-lo-MAX-comment-loc-per-doc}{320}
\DefMacro{corpus-lo-MIN-comment-loc-per-doc}{0}
\DefMacro{corpus-lo-MEDIAN-comment-loc-per-doc}{13.00}
\DefMacro{corpus-lo-STDEV-comment-loc-per-doc}{39.29}
\DefMacro{corpus-lo-AVG-total-loc-per-doc}{543.56}
\DefMacro{corpus-lo-SUM-total-loc-per-doc}{44,028}
\DefMacro{corpus-lo-MAX-total-loc-per-doc}{4,548}
\DefMacro{corpus-lo-MIN-total-loc-per-doc}{33}
\DefMacro{corpus-lo-MEDIAN-total-loc-per-doc}{342.00}
\DefMacro{corpus-lo-STDEV-total-loc-per-doc}{625.20}
\DefMacro{corpus-lo-AVG-code-loc-per-doc}{522.16}
\DefMacro{corpus-lo-SUM-code-loc-per-doc}{42,295}
\DefMacro{corpus-lo-MAX-code-loc-per-doc}{4,228}
\DefMacro{corpus-lo-MIN-code-loc-per-doc}{31}
\DefMacro{corpus-lo-MEDIAN-code-loc-per-doc}{327.00}
\DefMacro{corpus-lo-STDEV-code-loc-per-doc}{591.23}
\DefMacro{corpus-ta-num-projects}{20}
\DefMacro{corpus-ta-AVG-num-docs}{18.40}
\DefMacro{corpus-ta-AVG-spec-loc}{5,122.35}
\DefMacro{corpus-ta-AVG-proof-loc}{7,625.00}
\DefMacro{corpus-ta-AVG-comment-loc}{884.35}
\DefMacro{corpus-ta-AVG-total-loc}{13,631.70}
\DefMacro{corpus-ta-AVG-code-loc}{12,747.35}
\DefMacro{corpus-ta-SUM-num-docs}{368}
\DefMacro{corpus-ta-SUM-spec-loc}{102,447}
\DefMacro{corpus-ta-SUM-k-spec-loc}{102}
\DefMacro{corpus-ta-SUM-proof-loc}{152,500}
\DefMacro{corpus-ta-SUM-k-proof-loc}{152}
\DefMacro{corpus-ta-SUM-comment-loc}{17,687}
\DefMacro{corpus-ta-SUM-k-comment-loc}{17}
\DefMacro{corpus-ta-SUM-total-loc}{272,634}
\DefMacro{corpus-ta-SUM-k-total-loc}{272}
\DefMacro{corpus-ta-SUM-code-loc}{254,947}
\DefMacro{corpus-ta-SUM-k-code-loc}{254}
\DefMacro{corpus-ta-MAX-num-docs}{89}
\DefMacro{corpus-ta-MAX-spec-loc}{38,243}
\DefMacro{corpus-ta-MAX-k-spec-loc}{38}
\DefMacro{corpus-ta-MAX-proof-loc}{46,470}
\DefMacro{corpus-ta-MAX-k-proof-loc}{46}
\DefMacro{corpus-ta-MAX-comment-loc}{6,131}
\DefMacro{corpus-ta-MAX-k-comment-loc}{6}
\DefMacro{corpus-ta-MAX-total-loc}{90,844}
\DefMacro{corpus-ta-MAX-k-total-loc}{90}
\DefMacro{corpus-ta-MAX-code-loc}{84,713}
\DefMacro{corpus-ta-MAX-k-code-loc}{84}
\DefMacro{corpus-ta-MIN-num-docs}{1}
\DefMacro{corpus-ta-MIN-spec-loc}{70}
\DefMacro{corpus-ta-MIN-k-spec-loc}{0}
\DefMacro{corpus-ta-MIN-proof-loc}{8}
\DefMacro{corpus-ta-MIN-k-proof-loc}{0}
\DefMacro{corpus-ta-MIN-comment-loc}{16}
\DefMacro{corpus-ta-MIN-k-comment-loc}{0}
\DefMacro{corpus-ta-MIN-total-loc}{94}
\DefMacro{corpus-ta-MIN-k-total-loc}{0}
\DefMacro{corpus-ta-MIN-code-loc}{78}
\DefMacro{corpus-ta-MIN-k-code-loc}{0}
\DefMacro{corpus-ta-MEDIAN-num-docs}{12.50}
\DefMacro{corpus-ta-MEDIAN-spec-loc}{3,117.50}
\DefMacro{corpus-ta-MEDIAN-proof-loc}{3,368.00}
\DefMacro{corpus-ta-MEDIAN-comment-loc}{375.00}
\DefMacro{corpus-ta-MEDIAN-total-loc}{6,472.00}
\DefMacro{corpus-ta-MEDIAN-code-loc}{6,120.00}
\DefMacro{corpus-ta-STDEV-num-docs}{20.67}
\DefMacro{corpus-ta-STDEV-spec-loc}{8,100.97}
\DefMacro{corpus-ta-STDEV-proof-loc}{11,399.89}
\DefMacro{corpus-ta-STDEV-comment-loc}{1,399.00}
\DefMacro{corpus-ta-STDEV-total-loc}{20,541.67}
\DefMacro{corpus-ta-STDEV-code-loc}{19,157.05}
\DefMacro{corpus-ta-AVG-spec-loc-per-doc}{278.39}
\DefMacro{corpus-ta-SUM-spec-loc-per-doc}{102,447}
\DefMacro{corpus-ta-MAX-spec-loc-per-doc}{4,331}
\DefMacro{corpus-ta-MIN-spec-loc-per-doc}{1}
\DefMacro{corpus-ta-MEDIAN-spec-loc-per-doc}{159.00}
\DefMacro{corpus-ta-STDEV-spec-loc-per-doc}{378.53}
\DefMacro{corpus-ta-AVG-proof-loc-per-doc}{414.40}
\DefMacro{corpus-ta-SUM-proof-loc-per-doc}{152,500}
\DefMacro{corpus-ta-MAX-proof-loc-per-doc}{5,551}
\DefMacro{corpus-ta-MIN-proof-loc-per-doc}{0}
\DefMacro{corpus-ta-MEDIAN-proof-loc-per-doc}{251.00}
\DefMacro{corpus-ta-STDEV-proof-loc-per-doc}{577.02}
\DefMacro{corpus-ta-AVG-comment-loc-per-doc}{48.06}
\DefMacro{corpus-ta-SUM-comment-loc-per-doc}{17,687}
\DefMacro{corpus-ta-MAX-comment-loc-per-doc}{588}
\DefMacro{corpus-ta-MIN-comment-loc-per-doc}{0}
\DefMacro{corpus-ta-MEDIAN-comment-loc-per-doc}{32.00}
\DefMacro{corpus-ta-STDEV-comment-loc-per-doc}{58.76}
\DefMacro{corpus-ta-AVG-total-loc-per-doc}{740.85}
\DefMacro{corpus-ta-SUM-total-loc-per-doc}{272,634}
\DefMacro{corpus-ta-MAX-total-loc-per-doc}{5,619}
\DefMacro{corpus-ta-MIN-total-loc-per-doc}{1}
\DefMacro{corpus-ta-MEDIAN-total-loc-per-doc}{481.50}
\DefMacro{corpus-ta-STDEV-total-loc-per-doc}{827.33}
\DefMacro{corpus-ta-AVG-code-loc-per-doc}{692.79}
\DefMacro{corpus-ta-SUM-code-loc-per-doc}{254,947}
\DefMacro{corpus-ta-MAX-code-loc-per-doc}{5,560}
\DefMacro{corpus-ta-MIN-code-loc-per-doc}{1}
\DefMacro{corpus-ta-MEDIAN-code-loc-per-doc}{445.50}
\DefMacro{corpus-ta-STDEV-code-loc-per-doc}{790.57}
\DefMacro{corpus-allgroup-num-projects}{21}
\DefMacro{corpus-allgroup-AVG-num-docs}{21.38}
\DefMacro{corpus-allgroup-AVG-spec-loc}{5,474.48}
\DefMacro{corpus-allgroup-AVG-proof-loc}{8,679.90}
\DefMacro{corpus-allgroup-AVG-comment-loc}{924.76}
\DefMacro{corpus-allgroup-AVG-total-loc}{15,079.14}
\DefMacro{corpus-allgroup-AVG-code-loc}{14,154.38}
\DefMacro{corpus-allgroup-SUM-num-docs}{449}
\DefMacro{corpus-allgroup-SUM-spec-loc}{114,964}
\DefMacro{corpus-allgroup-SUM-k-spec-loc}{114}
\DefMacro{corpus-allgroup-SUM-proof-loc}{182,278}
\DefMacro{corpus-allgroup-SUM-k-proof-loc}{182}
\DefMacro{corpus-allgroup-SUM-comment-loc}{19,420}
\DefMacro{corpus-allgroup-SUM-k-comment-loc}{19}
\DefMacro{corpus-allgroup-SUM-total-loc}{316,662}
\DefMacro{corpus-allgroup-SUM-k-total-loc}{316}
\DefMacro{corpus-allgroup-SUM-code-loc}{297,242}
\DefMacro{corpus-allgroup-SUM-k-code-loc}{297}
\DefMacro{corpus-allgroup-MAX-num-docs}{89}
\DefMacro{corpus-allgroup-MAX-spec-loc}{38,243}
\DefMacro{corpus-allgroup-MAX-k-spec-loc}{38}
\DefMacro{corpus-allgroup-MAX-proof-loc}{46,470}
\DefMacro{corpus-allgroup-MAX-k-proof-loc}{46}
\DefMacro{corpus-allgroup-MAX-comment-loc}{6,131}
\DefMacro{corpus-allgroup-MAX-k-comment-loc}{6}
\DefMacro{corpus-allgroup-MAX-total-loc}{90,844}
\DefMacro{corpus-allgroup-MAX-k-total-loc}{90}
\DefMacro{corpus-allgroup-MAX-code-loc}{84,713}
\DefMacro{corpus-allgroup-MAX-k-code-loc}{84}
\DefMacro{corpus-allgroup-MIN-num-docs}{1}
\DefMacro{corpus-allgroup-MIN-spec-loc}{70}
\DefMacro{corpus-allgroup-MIN-k-spec-loc}{0}
\DefMacro{corpus-allgroup-MIN-proof-loc}{8}
\DefMacro{corpus-allgroup-MIN-k-proof-loc}{0}
\DefMacro{corpus-allgroup-MIN-comment-loc}{16}
\DefMacro{corpus-allgroup-MIN-k-comment-loc}{0}
\DefMacro{corpus-allgroup-MIN-total-loc}{94}
\DefMacro{corpus-allgroup-MIN-k-total-loc}{0}
\DefMacro{corpus-allgroup-MIN-code-loc}{78}
\DefMacro{corpus-allgroup-MIN-k-code-loc}{0}
\DefMacro{corpus-allgroup-MEDIAN-num-docs}{13.00}
\DefMacro{corpus-allgroup-MEDIAN-spec-loc}{3,298.00}
\DefMacro{corpus-allgroup-MEDIAN-proof-loc}{3,503.00}
\DefMacro{corpus-allgroup-MEDIAN-comment-loc}{394.00}
\DefMacro{corpus-allgroup-MEDIAN-total-loc}{6,689.00}
\DefMacro{corpus-allgroup-MEDIAN-code-loc}{6,451.00}
\DefMacro{corpus-allgroup-STDEV-num-docs}{24.18}
\DefMacro{corpus-allgroup-STDEV-spec-loc}{8,061.05}
\DefMacro{corpus-allgroup-STDEV-proof-loc}{12,084.10}
\DefMacro{corpus-allgroup-STDEV-comment-loc}{1,377.19}
\DefMacro{corpus-allgroup-STDEV-total-loc}{21,065.82}
\DefMacro{corpus-allgroup-STDEV-code-loc}{19,725.91}
\DefMacro{corpus-allgroup-AVG-spec-loc-per-doc}{256.04}
\DefMacro{corpus-allgroup-SUM-spec-loc-per-doc}{114,964}
\DefMacro{corpus-allgroup-MAX-spec-loc-per-doc}{4,331}
\DefMacro{corpus-allgroup-MIN-spec-loc-per-doc}{1}
\DefMacro{corpus-allgroup-MEDIAN-spec-loc-per-doc}{151.00}
\DefMacro{corpus-allgroup-STDEV-spec-loc-per-doc}{354.48}
\DefMacro{corpus-allgroup-AVG-proof-loc-per-doc}{405.96}
\DefMacro{corpus-allgroup-SUM-proof-loc-per-doc}{182,278}
\DefMacro{corpus-allgroup-MAX-proof-loc-per-doc}{5,551}
\DefMacro{corpus-allgroup-MIN-proof-loc-per-doc}{0}
\DefMacro{corpus-allgroup-MEDIAN-proof-loc-per-doc}{247.00}
\DefMacro{corpus-allgroup-STDEV-proof-loc-per-doc}{553.74}
\DefMacro{corpus-allgroup-AVG-comment-loc-per-doc}{43.25}
\DefMacro{corpus-allgroup-SUM-comment-loc-per-doc}{19,420}
\DefMacro{corpus-allgroup-MAX-comment-loc-per-doc}{588}
\DefMacro{corpus-allgroup-MIN-comment-loc-per-doc}{0}
\DefMacro{corpus-allgroup-MEDIAN-comment-loc-per-doc}{26.00}
\DefMacro{corpus-allgroup-STDEV-comment-loc-per-doc}{56.69}
\DefMacro{corpus-allgroup-AVG-total-loc-per-doc}{705.26}
\DefMacro{corpus-allgroup-SUM-total-loc-per-doc}{316,662}
\DefMacro{corpus-allgroup-MAX-total-loc-per-doc}{5,619}
\DefMacro{corpus-allgroup-MIN-total-loc-per-doc}{1}
\DefMacro{corpus-allgroup-MEDIAN-total-loc-per-doc}{462.00}
\DefMacro{corpus-allgroup-STDEV-total-loc-per-doc}{798.29}
\DefMacro{corpus-allgroup-AVG-code-loc-per-doc}{662.01}
\DefMacro{corpus-allgroup-SUM-code-loc-per-doc}{297,242}
\DefMacro{corpus-allgroup-MAX-code-loc-per-doc}{5,560}
\DefMacro{corpus-allgroup-MIN-code-loc-per-doc}{1}
\DefMacro{corpus-allgroup-MEDIAN-code-loc-per-doc}{428.00}
\DefMacro{corpus-allgroup-STDEV-code-loc-per-doc}{761.32}

\input{tables/numbers-lemma-metrics-filtered}
\input{tables/numbers-results-ln-t1--t1--t1}

\newcommand{\CorpusKLOC}{\UseMacro{corpus-t1-SUM-k-code-loc}k\xspace}
\newcommand{\CorpusNumProjects}{\numtoword{\UseMacro{corpus-t1-num-projects}}\xspace}

\newcommand{\YouTubeURL}{\url{https://youtu.be/HZ5ac7Q14rc}}

\newcommand{\CLIExampleURL}{\url{https://tinyurl.com/yy6l8fbq}}  %

\newcommand{\Pmathcomp}{math-comp\xspace}
\newcommand{\Poddorder}{odd-order\xspace}
\newcommand{\Pfourcolor}{fourcolor\xspace}
\newcommand{\Pfinmap}{finmap\xspace}

\newcommand{\PTEXTfcslpcm}{FCSL PCM\xspace}

\begin{document}

\bstctlcite{IEEEexample:BSTcontrol}

\title{\Title}

\author{
  \IEEEauthorblockN{
    Pengyu Nie\IEEEauthorrefmark{1},
    Karl Palmskog\IEEEauthorrefmark{2},
    Junyi Jessy Li\IEEEauthorrefmark{1},
    Milos Gligoric\IEEEauthorrefmark{1}
  }
  \IEEEauthorblockA{
    pynie@utexas.edu, palmskog@kth.se, jessy@austin.utexas.edu, gligoric@utexas.edu\\
    \makecell[c]{
      \IEEEauthorrefmark{1}
      \textit{The University of Texas at Austin},
      Austin, TX, USA
    }
    \qquad
    \makecell[c]{
      \IEEEauthorrefmark{2}
      \textit{KTH Royal Institute of Technology},
      Stockholm, Sweden
    }
  }
}

\maketitle
\thispagestyle{plain}
\pagestyle{plain}

\begin{abstract}
Naming conventions are an important concern in large verification projects
using proof assistants, such as Coq. In particular, lemma names are used by
proof engineers to effectively understand and modify Coq code. However, providing accurate and
informative lemma names is a complex task, which is currently often carried out
manually. Even when lemma naming is automated using rule-based tools,
generated names may fail to adhere to important conventions not specified explicitly.
We demonstrate a \tooltype, dubbed \CoqConvTool, which automatically
suggests lemma names in \Coq projects.
\CoqConvTool leverages a \nnmodel trained on existing \Coq code, thus avoiding
manual specification of naming conventions. To allow proof engineers to
conveniently access suggestions from \CoqConvTool during Coq project
development, we integrated the \tooltype into the popular Visual Studio Code editor.
Our evaluation shows that \CoqConvTool substantially outperforms
strong baselines for suggesting lemma names and is useful in practice.
The demo video for \CoqConvTool can be viewed at: \YouTubeURL.
\end{abstract}


\section{Introduction}
\label{sec:intro}

In large software projects with many contributors, names
of methods and classes are important for code
comprehension and modification. Open source projects often document
their naming conventions carefully, impose them on proposed
contributions, and willingly accept naming
fixes~\cite{AllamanisETAL14Learning}.

The \Coq proof assistant~\cite{Coq810} is increasingly used to develop
trustworthy software systems, e.g., compilers~\cite{Leroy2009} and
distributed systems\cite{Sergey2018}. As such verification projects grow in
scope and size, naming conventions become an important concern.
In particular, proof engineers use \emph{lemma names} to
effectively understand and modify code~\cite{Aspinall2016b}.

In contrast to method names in Java-like languages, which tend to use camel case and
regular English words (e.g., \CoqIn{openServerConnection}), \Coq
lemma names often mix camel case and underscores with heavily
abbreviated terminology from logic and advanced mathematics, which makes the naming
task more difficult. For example, in the Mathematical Components (MathComp) Coq library, the
lemma name \CoqIn{extprod_mulgA} is used to express
``\underline{a}ssociativity of \underline{mul}tiplication operations
in \underline{ext}ernal \underline{prod}uct \underline{g}roups'',
i.e., a property of abstract algebra. This meaning is obtained by
first decomposing the name into \CoqIn{extprod},
\CoqIn{mul}, \CoqIn{g}, and \CoqIn{A}, and then consulting the
MathComp naming conventions~\cite{MathCompContribGuide}.

Currently, documentation and enforcement of lemma naming conventions
in \Coq projects is largely a manual process. While some aspects of
naming conventions can be captured by rule-based tools, specification
of rules is tedious and often \emph{incomplete}. Moreover, most large
\Coq projects use mutually incompatible lemma naming schemes.

We present \CoqConvTool, a \tooltype which automatically suggests \Coq
lemma names. \CoqConvTool learns
naming conventions by leveraging neural networks trained on existing
\Coq code. The deep learning and suggestion processes use multiple
representations of lemma statements, including \strees and \Coq
\emph{\ktrees} (also called \emph{elaborated
  terms})~\cite{NieETAL20Deep}.
In essence, \CoqConvTool consists of (1) a set of components written
in OCaml that interact with Coq or directly process information
extracted from Coq, and (2) a set of components written in Python that
perform name learning and generation. The first set of components is
based on the SerAPI library~\cite{Gallego2016} for serialization of
Coq data, while the second set of components is based on the PyTorch
deep learning framework~\cite{PaszkeETAL17Automatic} and the OpenNMT
library~\cite{KleinETAL17OpenNMT}.

The core of \CoqConvTool is command-line based. Although valuable,
this does not provide a convenient interface for proof engineers as
they are stating and proving new Coq lemmas. Hence, we integrated the
\tooltype into Visual Studio Code (\VSCode)~\cite{VSCodeWebpage}, a
popular editor for Coq source code.

We evaluated an earlier version of \CoqConvTool using a corpus derived
from the \MathComp family of \Coq projects, finding that the \tooltype
significantly outperforms strong baselines on automatic
metrics~\cite{NieETAL20Deep}. Moreover, we found encouraging results
in a qualitative case study where the maintainer of a medium-sized Coq
project manually evaluated over 150 name suggestions generated by
\CoqConvTool.

The earlier \tooltype version provided few conveniences beyond basic
name suggestions via the command line, and did not include any editor integration. In addition
to the novel integration with \VSCode, the \tooltype version presented here provides
a significantly more automated installation process and supports system-level configuration
of Coq project and name suggestion parameters, making it suitable for
wider use by proof engineers.

Our code, documentation, and pre-trained models are publicly available
on GitHub: \\
\url{https://github.com/EngineeringSoftware/roosterize}.

\section{Technique and Implementation}
\label{sec:technique}

In this section, we explain the workflow of the \CoqConvTool
\tooltype, and then briefly describe our \nnmodel for lemma name
generation.

\subsection{\Tooltype Workflow}
\label{sec:technique:workflow}

\begin{figure}[t]
  \centering
  \resizebox{.9\columnwidth}{!}{{\large \begin{tikzpicture}
    \tikzset{node distance=3.5cm, every state/.style={thick},
      initial text={},
      double distance=2pt,
      every edge/.style={
        draw,
        color=black,
        >=stealth',
        auto,
        thick}
    }
    \tikzset{every state/.append style={rectangle,
        inner sep=10pt,
        minimum width = 3cm,
        text=black,
        draw=black,
        scale=0.52}
    }
    \tikzset{anno/.style={scale=0.5}}
    \node[state, fill=white, minimum height=4cm, rounded corners] (mutator) {\makecell{\\\\PyTorch}};
    \node[state, fill=white, above of=mutator, node distance=2cm, minimum width=2.5cm, inner sep=3pt, minimum height=1cm, rounded corners] (parser) {\tSubTokenizer};
    \node [right = 0.1cm of parser.north west] [anno] {\circled{3}};
    \node[state, fill=white, below of=parser, node distance=1.3cm, minimum width=2.5cm, inner sep=3pt, minimum height=0.9cm, rounded corners] (transformer) {\tModel};
    \node [right = 0.1cm of transformer.north west] [anno] {\circled{4}};
    \node[state, fill=white, below of=transformer, node distance=3cm, minimum width=2.6cm, inner sep=3pt, minimum height=0.9cm, rounded corners] (models) {\tDataMiner};
    \node [right = 0.1cm of models.north west] [anno] {\circled{2}};
    
    \node[state, fill=white, right of=parser, node distance=5cm, inner sep=5pt, minimum width=2.5cm, minimum height=1cm] (vfile) {Coq source/compiled files};
    \node[state, fill=white, below of=vfile, node distance=5.5cm, xshift=-1.5cm] (sexpfile) {syntax and terms};
    \node[state, fill=white, above of=sexpfile, node distance=2cm] (name) {lemma names};
    \node[state, fill=white, right of=mutator, node distance=10.5cm, minimum height=5cm, rounded corners] (coq) {\,\,\,Coq};
    \node[state, fill=white, right of=mutator, node distance=8.3cm, minimum height=2.3cm, rounded corners] (serapi) {\,\,\,\,\,\,\,\,\,\SerAPI};
    
    \node[state, fill=white, below of=vfile, node distance=1.9cm, xshift=0.4cm, rounded corners] (sercomp) {\texttt{ser}\{\texttt{comp},\texttt{tok},\texttt{name}\}};
    \node [right = 0.1cm of sercomp.north west] [anno] {\circled{1}};

    \draw[<-, >=stealth', auto, thick] (models.south) |- node[near start, right] {} (sexpfile.west);
    \draw[->] (vfile) edge node {} ([xshift=-0.2cm,yshift=0.3cm]sercomp);
    \draw[->, >=stealth', auto, thick] ([xshift=1.4cm]sercomp) |- node[below left] {} (sexpfile);
    \draw[->, >=stealth', auto, thick] (mutator.east) -| node[below right] {} ([xshift=-0.4cm]name.north);

\end{tikzpicture}}}
  \vspace{-2pt}
  \caption{Low-level workflow of \CoqConvTool.}
  \label{fig:workflow}
  \vspace{-4pt}
\end{figure}
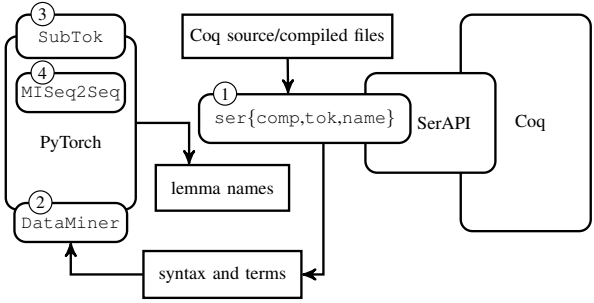

\figurename~\ref{fig:workflow} illustrates the low-level workflow of the \CoqConvTool
\tooltype. (1)~We use the \SerAPI library~\cite{Gallego2016} for extracting data from Coq
files, using three programs: \sertok for extracting
tokens, \sercomp for extracting the \strees, and \sername for
extracting \ktrees. \Strees are Coq's internal representations of
source code elements, including lemmas, during the parsing phase. \Ktrees
are Coq's internal representations of statements and functions during proof checking, and contain
rich information relevant for lemma naming. (2)~\tDataMiner orchestrates these
programs to obtain all lemmas in the given Coq files and their names, \lemmastmts,
and \sktrees.
(3)~\tSubTokenizer \subtokenize{s} the inputs for the \nnmodel.
(4)~\tModel is the multi-input \nnmodel for lemma name generation
(Section~\ref{sec:technique:model}). The model is implemented in the
popular deep learning framework PyTorch~\cite{PaszkeETAL17Automatic},
and is based on the OpenNMT library~\cite{KleinETAL17OpenNMT}.

Users interact with \CoqConvTool by using its \cli or the \VSCode
extension.  We use the Language Server Protocol (LSP)~\cite{LSPWebpage} to
connect the server (the core of \CoqConvTool including data extraction
scripts and the lemma name generation model) with the client (the
\VSCode extension). This simplifies future integration with other editors
that also support LSP, e.g., Emacs.

\subsection{\NNModel for Generating \LemmaN}
\label{sec:technique:model}

\begin{figure}[t]
  \centering
  \resizebox{0.8\columnwidth}{!}{\begin{tikzpicture}

  \tikzset{box/.style={
      text=white,
      draw=highlightcolor,
      fill=highlightcolor,
      font=\footnotesize,
  }}

  \tikzset{textbox/.style={
      draw=none,
      font=\footnotesize,
  }}

  \tikzset{textbox io/.style={
      draw=none,
      font=\scriptsize,
      text height=2mm,
      text depth=.4mm,
  }}

  \tikzset{nnode/.style={
      draw=black,
      minimum width=1mm,
      minimum height=1mm,
  }}

  \tikzset{anno/.style={
      font=\scriptsize,
  }}

  \node[coordinate] (c-anchor) at (0,0) {};
  \node[coordinate] (c-w-enc) at (-45mm,0) {};
  \node[coordinate] (c-w-dec) at (10mm,0) {};

  \node[coordinate] (c-w-enc1) [above = 14mm of c-w-enc] {};
  \node[coordinate] (c-w-enc2) at (c-w-enc) {};
  \node[coordinate] (c-w-enc3) [below = 14mm of c-w-enc] {};

  \newcommand{\wSepNNode}{4mm}

  \node[textbox](n-enc) [below right = 8mm and 6mm of c-w-enc3] {ENCODERS};
  \node[textbox] (n-in1) [above right = 2mm and -2mm of c-w-enc1] {\Light{lemma statement}};
  \node[nnode] (n-enc1-1) [right = 0 of c-w-enc1] {};
  \node[nnode] (n-enc1-2) [right = \wSepNNode of n-enc1-1] {};
  \node[nnode] (n-enc1-3) [right = \wSepNNode of n-enc1-2] {};
  \node[nnode] (n-enc1-4) [right = \wSepNNode of n-enc1-3] {};
  \node[textbox] (n-enc1-5) [right = \wSepNNode of n-enc1-4] {$\cdots$};
  \node[nnode] (n-enc1-6) [right = \wSepNNode of n-enc1-5] {};

  \foreach \i/\j in {1/2,2/3,3/4,4/5,5/6} {
    \draw[->] ([yshift=.3mm]n-enc1-\i.east) -- ([yshift=.3mm]n-enc1-\j.west);
    \draw[->] ([yshift=-.3mm]n-enc1-\j.west) -- ([yshift=-.3mm]n-enc1-\i.east);
  }

  \node[textbox io] (n-in1-1) [below = 2mm of n-enc1-1] {L};
  \node[textbox io] (n-in1-2) [below = 2mm of n-enc1-2] {1};
  \node[textbox io] (n-in1-3) [below = 2mm of n-enc1-3] {L};
  \node[textbox io] (n-in1-4) [below = 2mm of n-enc1-4] {2};
  \node[textbox io] (n-in1-6) [below = 2mm of n-enc1-6] {.};

  \foreach \i in {1,2,3,4,6} {
    \draw[->] (n-in1-\i) -- (n-enc1-\i);
  }

  \node[textbox] (n-in2) [above right = 2mm and -2mm of c-w-enc2] {\Light{chopped syntax tree}};
  \node[nnode] (n-enc2-1) [right = 0 of c-w-enc2] {};
  \node[nnode] (n-enc2-2) [right = \wSepNNode of n-enc2-1] {};
  \node[nnode] (n-enc2-3) [right = \wSepNNode of n-enc2-2] {};
  \node[nnode] (n-enc2-4) [right = \wSepNNode of n-enc2-3] {};
  \node[textbox] (n-enc2-5) [right = \wSepNNode of n-enc2-4] {$\cdots$};
  \node[nnode] (n-enc2-6) [right = \wSepNNode of n-enc2-5] {};

  \foreach \i/\j in {1/2,2/3,3/4,4/5,5/6} {
    \draw[->] ([yshift=.3mm]n-enc2-\i.east) -- ([yshift=.3mm]n-enc2-\j.west);
    \draw[->] ([yshift=-.3mm]n-enc2-\j.west) -- ([yshift=-.3mm]n-enc2-\i.east);
  }

  \node[textbox io] (n-in2-1) [below = 2mm of n-enc2-1] {(};
  \node[textbox io] (n-in2-2) [below = 2mm of n-enc2-2] {(};
  \node[textbox io] (n-in2-3) [below = 2mm of n-enc2-3] {(};
  \node[textbox io] (n-in2-4) [below = 2mm of n-enc2-4] {\ \ \ \ CLocalAssum};
  \node[textbox io] (n-in2-6) [below = 2mm of n-enc2-6] {)};

  \foreach \i in {1,2,3,4,6} {
    \draw[->] (n-in2-\i) -- (n-enc2-\i);
  }

  \node[textbox] (n-in3) [above right = 2mm and -2mm of c-w-enc3] {\Light{chopped kernel tree}};
  \node[nnode] (n-enc3-1) [right = 0 of c-w-enc3] {};
  \node[nnode] (n-enc3-2) [right = \wSepNNode of n-enc3-1] {};
  \node[nnode] (n-enc3-3) [right = \wSepNNode of n-enc3-2] {};
  \node[nnode] (n-enc3-4) [right = \wSepNNode of n-enc3-3] {};
  \node[textbox] (n-enc3-5) [right = \wSepNNode of n-enc3-4] {$\cdots$};
  \node[nnode] (n-enc3-6) [right = \wSepNNode of n-enc3-5] {};

  \foreach \i/\j in {1/2,2/3,3/4,4/5,5/6} {
    \draw[->] ([yshift=.3mm]n-enc3-\i.east) -- ([yshift=.3mm]n-enc3-\j.west);
    \draw[->] ([yshift=-.3mm]n-enc3-\j.west) -- ([yshift=-.3mm]n-enc3-\i.east);
  }

  \node[textbox io] (n-in3-1) [below = 2mm of n-enc3-1] {(};
  \node[textbox io] (n-in3-2) [below = 2mm of n-enc3-2] {Prod};
  \node[textbox io] (n-in3-3) [below = 2mm of n-enc3-3] {char};
  \node[textbox io] (n-in3-4) [below = 2mm of n-enc3-4] {(};
  \node[textbox io] (n-in3-6) [below = 2mm of n-enc3-6] {)};

  \foreach \i in {1,2,3,4,6} {
    \draw[->] (n-in3-\i) -- (n-enc3-\i);
  }

  \draw[draw=highlightcolor] ([yshift=-3mm]c-anchor) -- ([yshift=3mm]c-anchor) -- ([xshift=5mm]c-anchor) -- cycle;
  \node[anno] [below right = 5mm and -3mm of c-anchor] {\makecell[c]{fully\\connected\\layer}};
  \draw[->,rounded corners] (n-enc1-6) -- +(3mm,0) -- ([yshift=2mm]c-anchor);
  \draw[->,rounded corners] (n-enc2-6) -- +(3mm,0) -- ([yshift=0]c-anchor);
  \draw[->,rounded corners] (n-enc3-6) -- +(3mm,0) -- ([yshift=-2mm]c-anchor);

  \node[textbox] (n-out) [above right = 10mm and 8mm of c-w-dec] {\Light{lemma name}};
  \node[textbox] (n-dec) [below right = 10mm and 8mm of c-w-dec] {DECODER};

  \renewcommand{\wSepNNode}{4mm}

  \node[nnode] (n-dec-1) [right = 0 of c-w-dec] {};
  \node[nnode] (n-dec-2) [right = \wSepNNode of n-dec-1] {};
  \node[nnode] (n-dec-3) [right = \wSepNNode of n-dec-2] {};
  \node[nnode] (n-dec-4) [right = \wSepNNode of n-dec-3] {};
  \node[nnode] (n-dec-5) [right = \wSepNNode of n-dec-4] {};
  \node[nnode] (n-dec-6) [right = \wSepNNode of n-dec-5] {};

  \node[textbox io, font=\scriptsize] (n-out-0) [below = 2mm of n-dec-1] {$\langle$BOS$\rangle$};
  \node[textbox io] (n-out-1) [above = 2mm of n-dec-1] {mg};
  \node[textbox io] (n-out-2) [above = 2mm of n-dec-2] {\_};
  \node[textbox io] (n-out-3) [above = 2mm of n-dec-3] {eq};
  \node[textbox io] (n-out-4) [above = 2mm of n-dec-4] {\_};
  \node[textbox io] (n-out-5) [above = 2mm of n-dec-5] {nerode};
  \node[textbox io, font=\scriptsize] (n-out-6) [above = 2mm of n-dec-6] {\hspace{3mm}$\langle$EOS$\rangle$};

  \draw[->] ([xshift=5mm]c-anchor) -- (n-dec-1);
  \draw[->] (n-out-0) -- (n-dec-1);
  \draw[->] (n-dec-1) -- (n-out-1);
  \draw[->] (n-dec-2) -- (n-out-2);
  \draw[->] (n-dec-3) -- (n-out-3);
  \draw[->] (n-dec-4) -- (n-out-4);
  \draw[->] (n-dec-5) -- (n-out-5);
  \draw[->] (n-dec-6) -- (n-out-6);

  \draw[->] (n-dec-1.east) -- (n-dec-2.west);
  \draw[->] (n-dec-2.east) -- (n-dec-3.west);
  \draw[->] (n-dec-3.east) -- (n-dec-4.west);
  \draw[->] (n-dec-4.east) -- (n-dec-5.west);
  \draw[->] (n-dec-5.east) -- (n-dec-6.west);

  \draw[->] (n-out-1.east) -- ++(.5mm,0) -- ++(0,-9mm) -| (n-dec-2.south);
  \draw[->] (n-out-2.east) -- ++(1mm,0) -- ++(0,-9mm) -| (n-dec-3.south);
  \draw[->] (n-out-3.east) -- ++(.5mm,0) -- ++(0,-9mm) -| (n-dec-4.south);
  \draw[->] (n-out-4.east) -- ++(1mm,0) -- ++(0,-9mm) -| (n-dec-5.south);
  \draw[->] ([xshift=-1mm]n-out-5.east) -- ++(.5mm,0) -- ++(0,-9mm) -| (n-dec-6.south);

\end{tikzpicture}}
  \vspace{-2pt}
  \caption{Neural architecture of lemma name generation model in \CoqConvTool, exemplified for the name \CoqIn{mg_eq_nerode}.}
  \label{fig:model-architecture}
  \vspace{-4pt}
\end{figure}
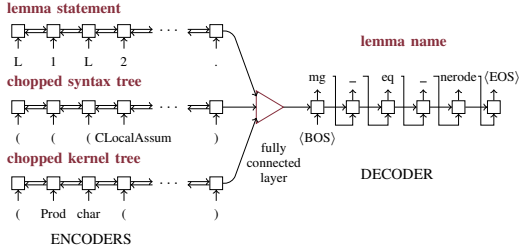

We consider lemma name generation with an \emph{\encdec} mindset, and
use sequence-to-sequence (\seqtoseq) neural architectures specifically designed for
transduction tasks~\cite{sutskever2014sequence}.
\figurename~\ref{fig:model-architecture} illustrates the architecture
of our model~\cite{NieETAL20Deep}.  The encoders are Recurrent Neural
Networks (RNNs) that learn a deep semantic representation of a given
\lemmastmt from its tokens, \stree, and \ktree.  The model can be
configured to use any combination of the three encoders.  The decoder
is another RNN that generates the descriptive lemma name based on the
input deep semantic representation.  We equipped the decoder RNN with
attention mechanism~\cite{LuongETAL15Effective} and copy
mechanism~\cite{SeeETAL17Get} to improve the generation accuracy.

All the inputs and the output are sequences of \subtok{s}; the
\subtok{s} of inputs are obtained using a \subtokenizer, and the
\subtok{s} of the output are concatenated to form the generated lemma
name.  We implemented the \subtokenizer based on the conventions
outlined by \MathComp developers~\cite{MathCompContribGuide} (e.g.,
the lemma name \CodeIn{extprod\_mulgA} should be \subtokenized to
\CodeIn{extprod}, \CodeIn{\_}, \CodeIn{mul}, \CodeIn{g}, and
\CodeIn{A}).
Because \sktrees can be large, we implemented \emph{\trimming}
heuristics to remove the parts irrelevant for generating \lemman
before feeding them to the encoders.  Our heuristics essentially:
(1)~replace the fully qualified name sub-trees with only the last
component of the name; (2)~remove the line number information from
sub-trees; (3)~extract the singletons, i.e., non-leaf nodes that have
only one child.

\section{Tool Installation}
\label{sec:tool-install}

\CoqConvTool currently supports macOS and Linux-based operating systems.
The first installation step is to download the \CoqConvTool repository:
\begin{lstlisting}[language=cmdline]
$\$$ git clone \
    (*@\ToolURL@*)
$\$$ cd roosterize (*@\&\&@*) git checkout v1.1.0+8.10.2
\end{lstlisting}

\MyPara{Required software and libraries} \CoqConvTool depends on two
sets of software and libraries: (1)~OCaml, \Coq, and \SerAPI;
(2)~PyTorch and other Python libraries.

To install OCaml (4.07.1), \Coq (8.10.2) and \SerAPI (0.7.1), we
recommend using the OCaml-based package-management system OPAM~\cite{OPAMWebpage}
version 2.0.7 or later:
\begin{lstlisting}[language=cmdline]
$\$$ opam switch create roosterize 4.07.1
$\$$ opam switch roosterize (*@\&\&@*) eval $\$$(opam env)
$\$$ opam update
$\$$ opam pin add coq 8.10.2
$\$$ opam pin add coq-serapi 8.10.0+0.7.1
\end{lstlisting}

To install PyTorch and other Python libraries, we recommend using the
package-management system Conda~\cite{CondaWebpage}.  The installation
script may be different depending on the operating system and whether
to use GPU or not.  For example, on Linux, to use CPU only:
\begin{lstlisting}[language=cmdline]
$\$$ conda env create --name roosterize \
                         --file conda-envs/cpu.yml
$\$$ conda activate roosterize
\end{lstlisting}

After installing these required software and libraries, users can use
\CoqConvTool via its \cli.

\MyPara{\VSCode extension} The \CoqConvTool \VSCode extension can be
installed easily from \VSCode marketplace: launch ``VS Code Quick
Open'' (Ctrl+P), paste the following command:
\begin{lstlisting}[language=cmdline]
ext install EngineeringSoftware.roosterize-vscode
\end{lstlisting}

Then, users should configure the path to \CoqConvTool executable file
(\CodeIn{./bin/roosterize}) using the following steps: open
``Settings'' (Ctrl+,), search for the entry ``\CoqConvTool: Bin
Path'', and fill in the path to \CoqConvTool executable file.

\begin{figure}[t]
  \centering
  \includegraphics[width=.85\columnwidth]{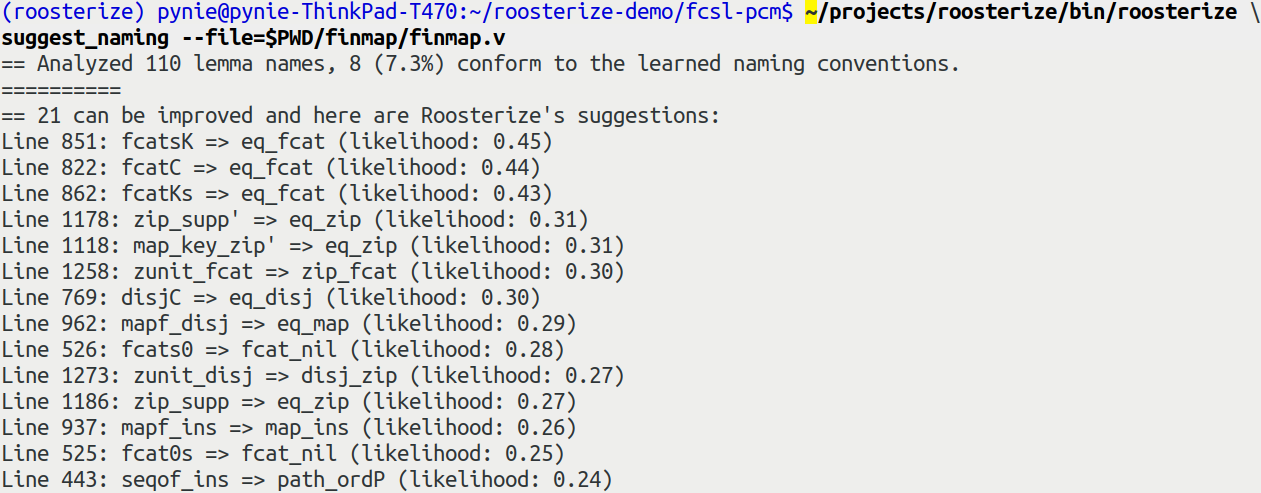}
  \caption{Screenshot of using \CoqConvTool from command line.}
  \label{fig:cli-suggestion}
  \vspace{-6pt}
\end{figure}

\begin{figure}[t]
  \centering
  \includegraphics[width=.7\columnwidth]{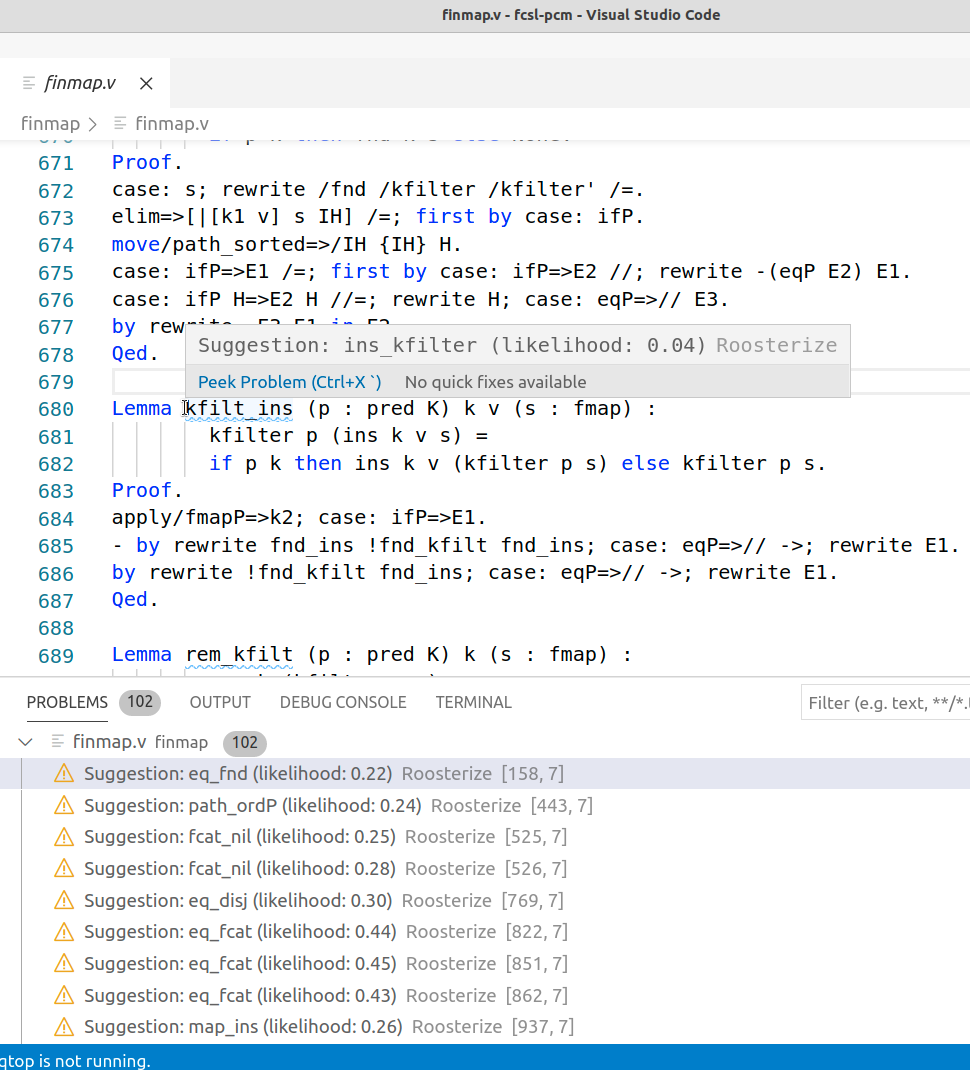}
  \caption{Screenshot of using \CoqConvTool from \VSCode.}
  \label{fig:vscode-suggestion}
  \vspace{-8pt}
\end{figure}

\section{Tool Usage}
\label{sec:tool-usage}

In this section, we describe the way our tool can be invoked from
command line and from \VSCode.

\subsection{Command Line}
\label{sec:tool-usage:cmdline}

After installation, users can launch \CoqConvTool via the executable
file \CodeIn{./bin/roosterize} (in short, \CodeIn{roosterize}).  We
focus on the main usage of \CoqConvTool---suggesting \lemman for Coq
projects. For other usages, e.g., training models, users can refer to
the help included in \CoqConvTool:
\begin{lstlisting}[language=cmdline]
$\$$ roosterize --help
\end{lstlisting}

Users should first obtain a model, e.g., by downloading a pre-trained
model.  The following command downloads the model we pre-trained on our \MathComp
corpus~\cite{NieETAL20Deep}:
\begin{lstlisting}[language=cmdline]
$\$$ roosterize download_global_model
\end{lstlisting}

Applying \CoqConvTool to a Coq project requires (1)~a
\CodeIn{\_CoqProject} file in the project root directory in the format used by
the \CodeIn{coq\_makefile} tool~\cite{CoqMakefile}, and (2)~that the project
source code has been compiled.
If a user specified the compilation command in the
\CodeIn{.roosterizerc} configuration file at the root directory of the
project, \CoqConvTool will automatically compile the project before
suggesting \lemman.  \CoqConvTool can suggest \lemman for one Coq \doc
at a time.  For example, running the following commands downloads the
Coq project \PTEXTfcslpcm at Git revision
\UseMacro{corpus-fcslpcm-sha-pretty}, prepares a
\CodeIn{.roosterizerc} configuration file, and suggests \lemman for
the \CodeIn{finmap.v} file in the project:
\begin{lstlisting}[language=cmdline]
$\$$ git clone \
    (*@\url{\UseMacro{corpus-fcslpcm-url}}@*)
$\$$ git checkout (*@\UseMacro{corpus-fcslpcm-sha-pretty}@*)
$\$$ echo "compile_cmd: make -j8" > ./.roosterizerc
$\$$ roosterize suggest_naming \
                 --file=$\$$PWD/finmap/finmap.v
\end{lstlisting}

In the last command, \CoqConvTool uses SerAPI to parse
\CodeIn{finmap.v} and extracts all lemmas, then uses the \lemman
suggestion model to generate top $k$ (default $k=5$) likely \lemman
for each lemma, and then compares the generated \lemman with the
original \lemman, and finally prints a report to suggest potential
\lemman changes.  \figurename~\ref{fig:cli-suggestion} shows the top
part of the report generated for \CodeIn{finmap.v} (full report available
at: \CLIExampleURL).

\subsection{\VSCode}
\label{sec:tool-usage:vscode}

Users should first open the Coq \docs they want to analyze. The steps
to obtain the \lemman suggestions for them are: (1)~open ``Command
Palettes'' (Ctrl+Shift+P), and (2)~choose
``Roosterize: Suggest Naming (for all .v files)''.
After \CoqConvTool produces the suggestions, the \lemman that do not
conform to the conventions are underlined, and users can hover the
mouse pointer over that underlined name to view \CoqConvTool's
suggestion in a tooltip.  Users can also view all suggestions in the
``Problems'' tab.  \figurename~\ref{fig:vscode-suggestion} shows a
screenshot of this step.

\section{Evaluation}
\label{sec:eval}

\subsection{Quantitative}

\MyPara{\Coq projects under study}
We selected \CorpusNumProjects large Coq projects from the \MathComp
family: \Pmathcomp, \Pfinmap, \Pfourcolor, and \Poddorder.  The
projects have a total of \CorpusKLOC{} lines of code,
\UseMacro{corpus-t1-SUM-num-docs} \docs, and
\UseMacro{lemma-filtered-t1-project-SUM-num-lemmas} lemmas.  More
information of these projects can be found in our
corpus~\cite{MathCompCorpus} (``Tier 1'' part).  Following standard
deep learning practice, we randomly split the projects' \docs into
training, validation, and testing sets which contain 80\%, 10\%, 10\%
of the \docs, which is \UseMacro{lemma-filtered-t1-train-num-lemmas},
\UseMacro{lemma-filtered-t1-val-num-lemmas},
\UseMacro{lemma-filtered-t1-test-num-lemmas} lemmas, respectively; the
lemmas from a same file are assigned to only one of the sets.

\MyPara{Results} We trained various configurations of our model.  In
this paper, we focus on five configurations that use different inputs:
\UseMacro{ln-s+bsexpl1+fsexpl1+attn+copy},
\UseMacro{ln-s+bsexpl1+attn+copy}, \UseMacro{ln-s+fsexpl1+attn+copy},
\UseMacro{ln-bsexpl1+fsexpl1+attn+copy}, and only
\UseMacro{ln-s+attn+copy} (where \LStmtAcro{} $=$ \lstmt,
\TrimmedSTreeAcro{} $=$ \trimmedstree,
\TrimmedKTreeAcro{} $=$ \trimmedktree).  We also compare our model with
a \retrievalbased baseline.

\begin{table}[t]
\begin{scriptsize}
\begin{center}
\vspace{5pt}
\caption{\UseMacro{table-caption-results-ln-t1--t1--t1-main}}
\vspace{-2pt}
\begin{tabular}{l r r r r}
\toprule
\TableHeadModel
& \UseMacro{table-head-results-ln-BLEU-4}
& \UseMacro{table-head-results-ln-frag-acc}
& \UseMacro{table-head-results-ln-full-acc-top-1}
& \UseMacro{table-head-results-ln-full-acc-top-5}
 \\
\midrule
\UseMacro{ln-s+bsexpl1+fsexpl1+attn+copy}
& 
\UseMacro{results-ln-t1--t1--t1-ln-s+bsexpl1+fsexpl1+attn+copy-test-AVG-BLEU-4}
& 
\UseMacro{results-ln-t1--t1--t1-ln-s+bsexpl1+fsexpl1+attn+copy-test-AVG-frag-acc}
& 
\UseMacro{results-ln-t1--t1--t1-ln-s+bsexpl1+fsexpl1+attn+copy-test-AVG-full-acc-top-1}
& 
\UseMacro{results-ln-t1--t1--t1-ln-s+bsexpl1+fsexpl1+attn+copy-test-AVG-full-acc-top-5}
\\
\UseMacro{ln-s+bsexpl1+attn+copy}
& 
\textbf{\UseMacro{results-ln-t1--t1--t1-ln-s+bsexpl1+attn+copy-test-AVG-BLEU-4}}
& 
\textbf{\UseMacro{results-ln-t1--t1--t1-ln-s+bsexpl1+attn+copy-test-AVG-frag-acc}}
& 
\textbf{\UseMacro{results-ln-t1--t1--t1-ln-s+bsexpl1+attn+copy-test-AVG-full-acc-top-1}}
& 
\textbf{\UseMacro{results-ln-t1--t1--t1-ln-s+bsexpl1+attn+copy-test-AVG-full-acc-top-5}}
\\
\UseMacro{ln-s+fsexpl1+attn+copy}
& 
\UseMacro{results-ln-t1--t1--t1-ln-s+fsexpl1+attn+copy-test-AVG-BLEU-4}
& 
\UseMacro{results-ln-t1--t1--t1-ln-s+fsexpl1+attn+copy-test-AVG-frag-acc}
& 
\UseMacro{results-ln-t1--t1--t1-ln-s+fsexpl1+attn+copy-test-AVG-full-acc-top-1}
& 
\UseMacro{results-ln-t1--t1--t1-ln-s+fsexpl1+attn+copy-test-AVG-full-acc-top-5}
\\
\UseMacro{ln-bsexpl1+fsexpl1+attn+copy}
& 
\UseMacro{results-ln-t1--t1--t1-ln-bsexpl1+fsexpl1+attn+copy-test-AVG-BLEU-4}
& 
\UseMacro{results-ln-t1--t1--t1-ln-bsexpl1+fsexpl1+attn+copy-test-AVG-frag-acc}
& 
\UseMacro{results-ln-t1--t1--t1-ln-bsexpl1+fsexpl1+attn+copy-test-AVG-full-acc-top-1}
& 
\UseMacro{results-ln-t1--t1--t1-ln-bsexpl1+fsexpl1+attn+copy-test-AVG-full-acc-top-5}
\\
\UseMacro{ln-s+attn+copy}
& 
\UseMacro{results-ln-t1--t1--t1-ln-s+attn+copy-test-AVG-BLEU-4}
& 
\UseMacro{results-ln-t1--t1--t1-ln-s+attn+copy-test-AVG-frag-acc}
& 
\UseMacro{results-ln-t1--t1--t1-ln-s+attn+copy-test-AVG-full-acc-top-1}
& 
\UseMacro{results-ln-t1--t1--t1-ln-s+attn+copy-test-AVG-full-acc-top-5}
\\
\midrule
\UseMacro{ln-rb}
& 
\UseMacro{results-ln-t1--t1--t1-ln-rb-test-AVG-BLEU-4}
& 
\UseMacro{results-ln-t1--t1--t1-ln-rb-test-AVG-frag-acc}
& 
\UseMacro{results-ln-t1--t1--t1-ln-rb-test-AVG-full-acc-top-1}
& 
\UseMacro{results-ln-t1--t1--t1-ln-rb-test-AVG-full-acc-top-5}
\\
\bottomrule
\end{tabular}
\end{center}
\end{scriptsize}
\vspace{\UseMacro{vspace-results-ln-t1--t1--t1-main}}
\end{table}

We evaluate each model by applying it on the testing set and measure
the average similarity between the generated names and the expected
names (as written by developers), using four automatic metrics:
\Bleu~\cite{PapineniETAL02BLEU}, \fragacc~\cite{NieETAL20Deep},
\toponeacc, and \topfiveacc.

Table~\ref{tbl:results-ln-t1--t1--t1-main} shows the results.  We
observed that \UseMacro{ln-s+bsexpl1+attn+copy} achieved the best
performance and substantially outperforms the \retrievalbased
baseline.  This shows the importance of using Coq's internal
structures. \Lstmt and \stree do not work well together primarily
because the two representations contain mostly the same information.
We performed extensive ablation studies to confirm the effectiveness
of the other parts of the model (Section 6.2 of our IJCAR'20
paper~\cite{NieETAL20Deep}), including the \trimming heuristics and
the attention and copy mechanisms.  We also performed a generalization
study which confirms that \CoqConvTool can perform well on a new
project with little additional training (Appendix D.3 of our IJCAR'20
paper~\cite{NieETAL20Deep}).

\subsection{Qualitative}

We carried out a qualitative case study using \CoqConvTool by applying
it to the \PTEXTfcslpcm Coq project, which comprises 690 lemmas. 36
suggestions (5\%) exactly matched the existing lemma names. We then
asked the project maintainer to comment on the remaining
suggestions. The maintainer found that 20\% of the suggested names he
inspected were of good quality, out of which more than half were of
high quality. Considering that the analysis was of top-1 suggestions,
we find these
results encouraging.

\section{Limitations and Future Work}
\label{sec:discussion}

Due to limitations in the protocol that \VSCode uses to communicate
with \Coq, our \VSCode extension cannot obtain name suggestions for a lemma
in real time as it is being edited,
i.e., by monitoring changes to the \Coq source file and proof
state.  However, our \tooltype can support this mode of use once
protocol limitations are lifted.

The quality of lemma name suggestions is highly dependent on the
quality of the pre-trained neural networks, and building a model
requires careful curation of Coq training data. While we have
constructed such a high-quality dataset based on the \MathComp family
of projects, additional datasets must be curated to suggest names that follow
conventions other than those for \MathComp.

\section{Conclusion}
\label{sec:conclusion}

We presented \CoqConvTool, a \tooltype for suggesting lemma names in
\Coq verification projects.
Nearly all related work addresses fundamentally different name
generation tasks in conventional languages such as Java~\cite{AllamanisETAL14Learning}.
An exception is Aspinall and Kaliszyk~\cite{Aspinall2016b}, who learn naming from a
corpus for the HOL Light proof assistant; however,
their technique only suggests names that appear in the training data.
\CoqConvTool uses novel neural network models pre-trained on existing
\Coq code to generate lemma names.
Our quantitative evaluation showed that \CoqConvTool outperforms
several strong baselines, and our qualitative evaluation demonstrated
the quality of generated lemma names.
We believe \CoqConvTool can be especially useful to proof engineers in
large \Coq projects to ensure that lemma names follow prevailing
conventions.  Through our integration of \CoqConvTool with the \VSCode
editor, naming suggestions can be continually provided as \Coq code is
added and revised.

\section*{Acknowledgments}

We thank Cyril Cohen, Emilio Jes{\'u}s Gallego Arias, Anton Trunov, and the
anonymous reviewers for their comments and feedback.
This work was partially supported by the US National Science
Foundation under Grant No. CCF-1652517 and the University of Texas at
Austin Continuing Fellowship.

\bibliographystyle{IEEEtran}
\bibliography{bib}

\end{document}